\documentclass[aps,prb,superscriptaddress,reprint,twocolumn,longbibliography]{revtex4-2}
\usepackage{bm}
\usepackage{amsmath,amssymb}
\usepackage{xcolor}
\usepackage{graphicx}
\usepackage[varg]{txfonts}

\usepackage[bookmarks=false,colorlinks=true,urlcolor=blue,citecolor=blue,linkcolor=blue,breaklinks=true]{hyperref}
\input glyphtounicode
\graphicspath{{figs/}}

\begin{document}
\title{Quantum-geometric shift of quasiequilibrium:\\Origin of nonreciprocal current driven by quantum-metric dipole}
\author{Sota Kitamura}
\affiliation{Department of Physics, Kyoto University, Kyoto, 606-8502, Japan}
\author{Takahiro Anan}
\affiliation{Department of Applied Physics, The University of Tokyo, Hongo, Tokyo,
  113-8656, Japan}
\affiliation{Department of Physics, Kyoto University, Kyoto, 606-8502, Japan}
\author{Takahiro Morimoto}
\affiliation{Department of Physics, Kyoto University, Kyoto, 606-8502, Japan}
\date{\today}
\begin{abstract}
We study nonlinear DC electric transport of quantum-metric origin by combining adiabatic perturbation theory with the nonequilibrium Green function approach.
The adiabatic ansatz provides a basis for directly treating a DC electric field in the velocity gauge, rather than introducing it as the zero-frequency limit of an AC field.
The resulting adiabatic-basis Hamiltonian takes the same form as in the length gauge, enabling a systematic comparison across different formulations.
Applying this fully quantum formulation, we find a longitudinal nonreciprocal current governed by the quantum-metric dipole.
The essential ingredient is a quantum correction to the distribution function that is absent in semiclassical treatments.
We trace this correction to the finite spread of an electron wave packet during relaxation under a bias field, thereby identifying shifted quasiequilibrium as the physical origin of quantum-metric nonreciprocal transport.
\end{abstract}
\maketitle

\section{Introduction}

The geometric and topological characterization of electronic states has become 
a central paradigm in modern condensed matter physics~\cite{Xiao2010,Nagaosa2010,Hasan2010,Qi2011}. 
Geometry and topology play a central role in various equilibrium and linear-response phenomena, 
ranging from electric polarization~\cite{KingSmith1993,Resta1994} and anomalous Hall transport~\cite{Nagaosa2010} to topological phases of matter~\cite{Hasan2010,Qi2011}. 
The manifestation of wave-function geometry in response phenomena is closely tied to 
the fact that external fields drive Bloch electrons through momentum space, 
causing the associated wave functions to evolve on a geometrically nontrivial manifold. 
Geometric quantities such as the Berry connection and curvature naturally 
emerge as signatures of this evolution~\cite{Berry1984,Xiao2010}. 
Geometric response is thus rooted in field-driven nonequilibrium dynamics, 
with linear response representing merely its leading-order manifestation. 
From this perspective, nonlinear transport offers a unique opportunity to uncover richer aspects of wave-function geometry.

Indeed, nonlinear response phenomena have attracted considerable interest in recent years~\cite{Tokura2018,Morimoto2023,Nagaosa2024,SurezRodrguez2025}. A wide variety of effects have been investigated, ranging from optical responses such as shift and injection currents~\cite{Sipe2000,Morimoto2016} and high-harmonic generation~\cite{Aversa1995,Ghimire2019} to DC transport phenomena including nonlinear Hall effects~\cite{Sodemann2015,Ma2019,Kang2019,Gao2023,Han2024} and nonreciprocal charge transport~\cite{Rikken2001,Wakatsuki2017,Tokura2018,Morimoto2018,Kitamura2020-2,Wang2023}. 
These studies revealed a crucial role of geometric quantities encoded in Bloch wave functions in nonlinear responses.

Among various geometric quantities, the quantum metric characterizes the distance structure of quantum states~\cite{Provost1980,Liu2025,Yu2025}. The quantum metric $g^{ij}_n$ is defined as the real part of the quantum geometric tensor $Q_n^{ij}=\langle\partial^i u_n|(1-|u_n\rangle\langle u_n|)|\partial^ju_n\rangle$ by
\begin{equation}
g_{n}^{ij}(\bm{k})=\mathrm{Re}\left[\frac{\partial\langle u_{n\bm{k}}|}{\partial k_i}(1-|u_{n\bm{k}}\rangle\langle u_{n\bm{k}}|)\frac{\partial |u_{n\bm{k}}\rangle}{\partial k_j}\right]
\end{equation}
for the Bloch state $|u_{n\bm{k}}\rangle$ of band $n$, momentum $\bm{k}$, thereby playing a complementary role to the Berry curvature $\Omega_{n}^{ij}=-2\mathrm{Im}\,Q_{n}^{ij}$.
Its physical significance was first recognized through the gauge-invariant part of the Wannier spread~\cite{Marzari1997}, and was later highlighted by the geometric contribution to the superfluid weight in flat-band superconductors~\cite{Peotta2015,Liang2017}. 
Quantum geometry has subsequently been shown to govern the stability of fractional Chern insulators~\cite{Roy2014,Jackson2015} and to constrain the energy gap and structure factors~\cite{Onishi2024-1,Onishi2024-2}, further broadening the scope of quantum geometry. 
These developments have stimulated increasing interest in its possible manifestations in nonlinear responses.

Among nonlinear responses, DC electric transport has attracted particular attention as a testbed for quantum geometry, as exemplified by the Berry-curvature dipole mechanism for the nonlinear Hall effect~\cite{Sodemann2015,Ma2019,Kang2019}. 
In view of such geometric roles of the Berry curvature ($\text{Im}Q$) in transverse responses, the quantum metric ($\text{Re}Q$) is expected to provide a related geometric mechanism in DC electric transport, especially for longitudinal responses.
A pioneering work in this direction is Ref.~\cite{Gao2014}, which showed that an electric field induces a correction to the Berry connection, characterized by
\begin{equation}
G^{ij}_n(\bm{k})=
\mathrm{Re}\sum_{m\neq n}\frac{\partial\langle u_{n\bm{k}}|}{\partial k_i}\frac{|u_{m\bm{k}}\rangle\langle u_{m\bm{k}}|}{\epsilon_n(\bm{k})-\epsilon_m(\bm{k})}\frac{\partial |u_{n\bm{k}}\rangle}{\partial k_j}
\end{equation}
with eigenenergy $\epsilon_n(\bm{k})$, 
which is referred to as the Berry-connection polarizability (BCP), with an overall factor of two.
Because of its formal resemblance to the quantum metric, $G^{ij}_n$ is sometimes called a band-normalized quantum metric.
Although this correction was originally shown to contribute to an intrinsic transverse response, it has inspired extensive studies on nonlinear DC conductivity in a variety of formulations, aiming at a geometric understanding of transverse and longitudinal DC responses~\cite{Watanabe2020,Oiwa2022,Michishita2022,Kaplan2023SciPost,Das2023,Wang2024,Kaplan2024}.
In particular, several works have focused on the role of metric-like quantities in longitudinal transport and highlighted a possible intrinsic contribution governed by the BCP~\cite{Das2023,Wang2024,Kaplan2024}.

These developments stimulated closer examination of the conditions under which an intrinsic longitudinal current can arise, including the issue of how it is sustained in a steady state.
If an intrinsic longitudinal current were present even in the clean limit with no dissipation channel, 
the associated power absorption $\bm{J}\cdot\bm{E}$ cannot be balanced in the steady state.
On this basis, some recent studies argue that it should vanish identically~\cite{Xiao2025,Qiang2026,Tang2026}. 
At the same time, another viewpoint is introduced by the recent 
proposal of a longitudinal response originating from the quantum metric $g^{ij}$ rather than from the BCP $G^{ij}$~\cite{Ulrich2026,Guo2026,Anan2026}.
This raises a further question of how such a metric-induced response can be reconciled with the issue of energy balance for the intrinsic longitudinal current.

Thus, despite the growing literature on nonlinear DC conductivity~\cite{Watanabe2020,Oiwa2022,Michishita2022,Das2023,Wang2024,Kaplan2024,Xiao2025,Qiang2026,Tang2026,Ulrich2026,Guo2026,Anan2026}, 
its unified understanding has not yet been reached. 
A central difficulty is that different formulations, including velocity- and length-gauge approaches, often yield different expressions and sometimes different conclusions for the nonlinear conductivity. 
The precise origin of this disagreement has remained difficult to pin down within the existing formulations.

\begin{figure*}
    \centering
    \includegraphics[width=.8\linewidth]{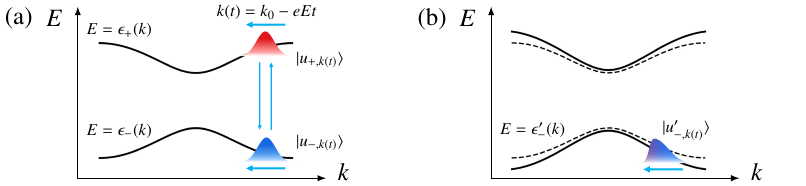}
    \caption{(a) Schematic picture for the adiabatic-basis representation. Each basis function depicted as a wave packet smoothly evolves along the equilibrium band structure, with a drift of momentum $k(t)=k_0-eEt$. Nonadiabatic transitions due to nonzero $E$ occur between different bands. 
    (b) Schematic picture for the field-modulated adiabatic wave function. By taking account of interband transitions between equilibrium bands perturbatively, one can redefine the adiabatic wave function and associated band structure as field-modulated quantities. Typically the first-order correction describes the Stark polarization, which is schematically represented by a skewed packet. The resultant basis function does not exhibit (perturbative) interband transition and provides a solution to the time-dependent Schr\"odinger equation.  }
    \label{fig:unitary}
\end{figure*}

In this study, we revisit the nonlinear conductivity of quantum-metric origin from the viewpoint of adiabatic perturbation theory~\cite{Berry1984,Rigolin2008,DeGrandi2010,Kitamura2020} applied to the velocity-gauge formalism.
In the velocity-gauge formalism, a DC electric field encoded by the time-linear vector potential $\bm{A}(t)=-\bm{E}t$ prevents a straightforward application of standard perturbation theory, usually requiring one to consider AC driving and take the low-frequency limit.
The adiabatic ansatz, depicted in Fig.~\ref{fig:unitary}(a), provides a solution to the time-dependent Schr\"odinger equation under slow parametric modulation,
and thus serves as a suitable basis for the direct treatment of DC electric fields in the velocity-gauge formulation.

In this adiabatic-basis representation, the Hamiltonian takes the same form as in the length gauge,
while avoiding the direct manipulation of the position operator. 
The adiabatic-basis representation therefore enables us to bridge the velocity- and length-gauge formulations of DC electric transport.
Using adiabatic perturbation theory, we construct the field-modulated wave functions [depicted in Fig.~\ref{fig:unitary}(b)] and confirm that they coincide with those used in the length-gauge formalism.
This implies that the discrepancy in the nonlinear current should originate from the assumption underlying the nonequilibrium distribution function. 

Then, using the nonequilibrium Green function approach, 
we determine the nonequilibrium distribution function for the wave functions employed in the length-gauge formalism 
with a fully quantum treatment. 
We obtain a nonlinear current identical to that reported in the velocity-gauge formalism~\cite{Ulrich2026,Guo2026,Anan2026}, 
whereas the same form has not been reproduced in the length-gauge formalism so far.
In particular, we obtain a longitudinal current
of quantum-metric origin, while that derived from the BCP vanishes identically.
We identify the origin of this metric-induced longitudinal transport
as a hidden quantum correction to the quasiequilibrium distribution, which is absent in the semiclassical treatment~\cite{Xiao2025,Qiang2026,Tang2026}. 
The adiabatic-basis representation compatible with the length-gauge picture enables us to interpret this quantum-metric correction 
as an inevitable inhomogeneity in the relaxation process under a bias field, reflecting the finite spread of the wave packet (See Fig.~\ref{fig:metric} for a schematic illustration).
Our finding highlights a novel type of geometric correction involving nonequilibrium relaxation processes.

The rest of this paper is organized as follows. 
In Sec.~\ref{sec:adiabatic}, we introduce the adiabatic-basis formalism, 
which reformulates the velocity-gauge problem in a basis adapted to the field-driven evolution and yields a Hamiltonian analogous to that in the length gauge without directly invoking the position operator. 
In Sec.~\ref{sec:current}, we combine this formalism with the nonequilibrium Green function approach to derive the nonlinear DC current and identify the quantum-metric contribution to the longitudinal current. 
In Sec.~\ref{sec:origin}, we clarify the physical origin of this correction to the quasiequilibrium distribution function in terms of the finite spread of the field-modulated wave packet during the relaxation process. 
In Sec.~\ref{sec:comparison}, we compare the different formalisms for the nonlinear DC conductivity in terms of the resulting density matrix and identify the origin of the different expressions.
Finally, we conclude the paper in Sec.~\ref{sec:conclusion} by discussing the distinction between intrinsic and relaxation-mediated nonlinear currents.

\section{The adiabatic-basis formalism}\label{sec:adiabatic}

In this section, we formulate the electron dynamics under an electric field, using the adiabatic-basis representation in the velocity gauge.
We consider the time-dependent problem described by the Hamiltonian
\begin{equation}
H(\bm{k}(t))=H(\bm{k}_{0}+e\bm{A}(t)),
\end{equation}
whose instantaneous eigenstates are defined as
\begin{equation}
H(\bm{k})|u_{n\bm{k}}\rangle=\epsilon_{n}(\bm{k})|u_{n\bm{k}}\rangle.
\end{equation}

When the speed of the parameter change, i.e. the electric field $\dot{\bm{k}}=e\dot{\bm{A}}(t)=-e\bm{E}(t)$,
is infinitesimally slow, the solution of the time-dependent Schr\"{o}dinger
equation 
\begin{equation}
i\frac{\partial}{\partial t}|\psi_{\bm{k}_{0}}(t)\rangle=H(\bm{k}(t))|\psi_{\bm{k}_{0}}(t)\rangle\label{eq:tdse}
\end{equation}
can be written as 
\begin{equation}
|\psi_{n\bm{k}_{0}}^{\text{ad}}(t)\rangle=|u_{n\bm{k}(t)}\rangle e^{-i\gamma_{n\bm{k}_{0}}(t)}\label{eq:ad}
\end{equation}
according to the adiabatic theorem~\cite{Berry1984,Rigolin2008}.
The phase factor $\gamma$ consists of the dynamical
and Berry phases, which are given by 
\begin{align}
\gamma_{n\bm{k}_{0}}(t) & =\int_{0}^{t}dt^{\prime}[\epsilon_{n}(\bm{k}(t^{\prime}))+e\bm{E}(t^{\prime})\cdot\bm{a}_{n}(\bm{k}(t^{\prime}))],\label{eq:gamma-ad}
\end{align}
with 
\begin{align}
\bm{a}_{n}(\bm{k}) & =\langle u_{n\bm{k}}|i\nabla_{\bm{k}}|u_{n\bm{k}}\rangle
\end{align}
being the Berry connection. This provides a minimal manifestation
of the geometric structure behind the Bloch wave function, where the
overall phase of the instantaneous eigenstate $|u_{n\bm{k}(t)}\rangle$
combined with the Berry-phase correction realizes a smooth deformation
of the wave function under $\bm{k}(t)=\bm{k}_{0}+e\bm{A}(t)$. 

We use this adiabatic time evolution as a basis function~\cite{Rigolin2008,DeGrandi2010,Kitamura2020},
and express the non-adiabatic evolution of the wave function
$|\psi_{\bm{k}_{0}}(t)\rangle$ described by Eq.~(\ref{eq:tdse})
as 
\begin{align}
|\psi^\prime_{m\bm{k}_{0}}(t)\rangle & =\sum_{n}C_{nm}(t)|u_{n\bm{k}(t)}\rangle e^{-i\gamma^\prime_{n\bm{k}_{0}}(t)}.\label{eq:psi}
\end{align}
Here we impose an initial condition at $t=-\infty$ as $C_{nm}(-\infty)=\delta_{nm}$,
at which the electric field is assumed to be absent and then adiabatically
switched on. 
In this basis, interband transitions are dictated by nonzero $C_{nm}$, which is schematically depicted in Fig.~\ref{fig:unitary}(a). As shown below, a perturbative construction of $C_{nm}(t)$ yields the form
$C_{nm}(t)=\tilde{C}_{nm}(\bm{k}(t))e^{i\gamma^\prime_{nm}(t)}$ with $\gamma^\prime_{nm}=\gamma^\prime_{n\bm{k}_{0}}-\gamma^\prime_{m\bm{k}_{0}}$,
so that the wave function reads
\begin{gather}
|\psi^\prime_{m\bm{k}_{0}}(t)\rangle=|u^\prime_{m\bm{k}(t)}\rangle e^{-i\gamma^\prime_{m\bm{k}_{0}}(t)},\label{eq:psi-2}\\
|u^\prime_{m\bm{k}(t)}\rangle=\sum_{n}\tilde{C}_{nm}(\bm{k}(t))|u_{n\bm{k}(t)}\rangle.
\end{gather}
This expression demonstrates that the time evolution under a finite electric field can still be written in the adiabatic form Eq.~(\ref{eq:ad}), but with the field-modulated instantaneous eigenstate $|u^\prime_{m\bm{k}}\rangle$ and the phase factor $\gamma^\prime_{m\bm{k}_{0}}$. We thus refer to $|\psi^\prime_{m\bm{k}_{0}}(t)\rangle$ as the field-modulated adiabatic wave function hereafter. This field-modulated adiabatic wave function undergoes a time evolution with no interband transition, which is depicted in Fig.~\ref{fig:unitary}(b). We note that the absence of interband transitions is ensured only at the perturbative level, and the nonperturbative contribution due to the Landau-Zener tunneling becomes relevant at high fields~\cite{Kitamura2020,Kitamura2020-2}.

By substituting Eq.~(\ref{eq:psi}) into Eq.~(\ref{eq:tdse}), we
obtain the equation of motion for $C_{nm}(t)=\langle u_{n\bm{k}(t)}|\psi^\prime_{m\bm{k}_{0}}(t)\rangle e^{i\gamma^\prime_{n\bm{k}_{0}}(t)}$
as 
\begin{align}
i\dot{C}_{nm}(t) & =\left[\epsilon_{n}(\bm{k}(t))+e\bm{E}(t)\cdot\bm{a}_{n}(\bm{k}(t))-\dot{\gamma}^\prime_{n\bm{k}_{0}}(t)\right]C_{nm}(t)\nonumber \\
 & +e\sum_{l\neq n}\bm{E}(t)\cdot\bm{\xi}_{nl}(\bm{k}(t))e^{i\gamma^\prime_{nl}(t)}C_{lm}(t),\label{eq:eom-C-0}
\end{align}
where we define the electric dipole moment $-e\bm{\xi}_{nm}$, as
\begin{align}
\bm{\xi}_{nm}(\bm{k}) & =\langle u_{n\bm{k}}|i\nabla_{\bm{k}}|u_{m\bm{k}}\rangle(1-\delta_{nm}).
\end{align}
Note that the matrix element in the right-hand side of Eq.~(\ref{eq:eom-C-0})
has the same form as the Hamiltonian often employed in the length-gauge
formalism, when we set $\gamma^\prime=0$ and $\bm{k}(t)\to\bm{k}_{0}$. 

We expand the phase factor $\gamma^\prime_{n\bm{k}_{0}}(t)$ as
\begin{equation}
\gamma^\prime_{n\bm{k}_{0}}(t)=\gamma_{n\bm{k}_{0}}^{(0)}(t)+\gamma_{n\bm{k}_{0}}^{(1)}(t)+\gamma_{n\bm{k}_{0}}^{(2)}(t)+\dots,
\end{equation}
 in order to eliminate secular terms in the calculation. Here the superscript indicates that $\dot{\gamma}^{(n)}$ for a given $\bm{k}(t)$ is $n$th order in $E$. $\gamma^\prime_{n\bm{k}_{0}}(t)$
is determined order by order such that the time-integral of the equation
of motion over $[-\infty,t]$ does not diverge. As evident from
Eq.~(\ref{eq:eom-C-0}), the zeroth and first order terms vanish
with 
\begin{align}
\dot{\gamma}_{n\bm{k}_{0}}^{(0)}(t) & =\epsilon_{n}(\bm{k}(t)),\label{eq:gamma-0}\\
\dot{\gamma}_{n\bm{k}_{0}}^{(1)}(t) & =e\bm{E}(t)\cdot\bm{a}_{n}(\bm{k}(t)),\label{eq:gamma-1}
\end{align}
as in the adiabatic limit Eq.~(\ref{eq:gamma-ad}). Then the equation
of motion is simplified into
\begin{align}
i\dot{C}_{nm}(t) & =-\dot{\gamma}_{n\bm{k}_{0}}^{(2)}(t)C_{nm}(t)\nonumber \\
 & +e\sum_{l\neq n}\bm{E}(t)\cdot\bm{\xi}_{nl}(\bm{k}(t))e^{i\gamma^\prime_{nl}(t)}C_{lm}(t).\label{eq:eom-C}
\end{align}

As we detail in Appendix~\ref{sec:appendix-derivation-C}, we can construct the formal solution by integrating this equation over $[-\infty,t]$
and performing a recursive substitution. While the obtained formal
solution is expressed in terms of the nested time integral (as in
the standard time-dependent perturbation), we can evaluate the integral
as a series in $E$ by using integration
by parts. This gives $C_{nm}(t)=\tilde{C}_{nm}(\bm{k}(t))e^{i\gamma^\prime_{nm}(t)}$
with 
\begin{align}
\tilde{C}_{nm}(\bm{k}) & =\left(1-\frac{1}{2}\sum_{l\neq n}\left|\frac{e\bm{E}\cdot\bm{\xi}_{nl}}{\Delta_{nl}}\right|^{2}\right)\delta_{nm}\nonumber \\
 & +\Biggl[-e\frac{\bm{E}\cdot\bm{\xi}_{nm}}{\Delta_{nm}}+\frac{ie^{2}}{\Delta_{nm}}\bm{E}\cdot\hat{\bm{D}}_{nm}\left(\frac{\bm{E}\cdot\bm{\xi}_{nm}}{\Delta_{nm}}\right)\nonumber \\
 & +e^{2}\sum_{l\neq n,m}\frac{\bm{E}\cdot\bm{\xi}_{nl}\bm{E}\cdot\bm{\xi}_{lm}}{\Delta_{lm}\Delta_{nm}}\Biggr](1-\delta_{nm})+O(E^{3})\label{eq:C}
\end{align}
for the DC electric field, 
where we define $\Delta_{nm}(\bm{k})=\epsilon_{n}(\bm{k})-\epsilon_{m}(\bm{k})$ and $\hat{\bm{D}}_{nm}(\bm{k})=\nabla_{\bm{k}}-i(\bm{a}_{n}(\bm{k})-\bm{a}_{m}(\bm{k}))$.
This formula for $\tilde{C}_{nm}$ describes the field modulation of the instantaneous eigenstate $|u_{m\bm{k}}^\prime\rangle$ up to $E^{2}$. We also obtain the
correction to the phase factor as 
\begin{align}
\dot{\gamma}_{n\bm{k}_{0}}^{(2)}(t) & =\sum_{m\neq n}\frac{|e\bm{E}\cdot\bm{\xi}_{nm}(\bm{k}(t))|^{2}}{\Delta_{nm}(\bm{k}(t))}.\label{eq:gamma-2}
\end{align}
The field-modulated wave function has the same form as the field-dressed Bloch states in the length-gauge formalism, while the perturbed eigenenergy is obtained as $\dot{\gamma}^\prime$ in the length gauge~\cite{Qiang2026,Tang2026}.

We can write down the matrix element
of the velocity operator $\nabla_{\bm{k}}H(\bm{k}(t))$ for $|\psi^\prime_{n\bm{k}_{0}}(t)\rangle$
defined as Eq.~(\ref{eq:psi}), as 
\begin{align}
\bm{V}^\prime_{nm}(t) & =\langle\psi^\prime_{n\bm{k}_{0}}(t)|\nabla_{\bm{k}}H(\bm{k}(t))|\psi^\prime_{m\bm{k}_{0}}(t)\rangle\nonumber \\
&=\frac{d}{dt}\langle\psi_{n\bm{k}_{0}}^{\prime}(t)|i\nabla_{\bm{k}}|\psi_{m\bm{k}_{0}}^{\prime}(t)\rangle \nonumber \\
 & =\bm{V}_{n}^{\prime\text{diag}}(\bm{k}(t))\delta_{nm}+\bm{V}_{nm}^{\prime\text{off}}(\bm{k}(t))e^{i\gamma_{nm}^{\prime}(t)}(1-\delta_{nm}),\label{eq:V}
\end{align}
where the diagonal component takes the form
\begin{align}
\bm{V}_{n}^{\prime\text{diag}}(\bm{k}) & =\nabla_{\bm{k}}\epsilon^\prime_{n}+e\bm{E}\times\left(\nabla_{\bm{k}}\times\bm{a}^\prime_{n}\right),\label{eq:Vdiag}
\end{align}
while the offdiagonal matrix element reads
\begin{align}
\bm{V}_{nm}^{\prime\text{off}}(\bm{k}) & =i\Delta_{nm}^{\prime}\bm{\xi}_{nm}^{\prime}-e\bm{E}\cdot\hat{\bm{D}}_{nm}^{\prime}\bm{\xi}_{nm}^{\prime}
\end{align}
to arbitrary order in $E$.
Here we have defined the field-modulated Berry connection $\bm{a}_{n}^{\prime}$ by
\begin{align}
\bm{a}^\prime_{n}(\bm{k}) & =\langle u^\prime_{n\bm{k}}|i\nabla_{\bm{k}}|u^\prime_{n\bm{k}}\rangle\label{eq:aprime-2},
\end{align}
and analogously the modulated electric dipole $-e\bm{\xi}_{nm}^\prime$.
The field-modulated instantaneous eigenenergy $\epsilon_{n}^{\prime}$ has then been introduced through
\begin{equation}
\dot{\gamma}_{n\bm{k}_{0}}^{\prime}(t)=\epsilon_{n}^{\prime}(\bm{k}(t))+e\bm{E}\cdot\bm{a}_{n}^{\prime}(\bm{k}(t)),
\end{equation}
along with $\Delta^\prime_{nm}=\epsilon^\prime_{n}-\epsilon^\prime_{m}$, $\hat{\bm{D}}^\prime_{nm}=\nabla_{\bm{k}}-i(\bm{a}^\prime_{n}-\bm{a}^\prime_{m})$.
The modulated eigenenergy can be shown to coincide
with 
\begin{align}
\epsilon^\prime_{n}(\bm{k}) & =\langle u^\prime_{n\bm{k}}|H(\bm{k})|u^\prime_{n\bm{k}}\rangle\label{eq:epsprime-h}
\end{align}
using Eq.~(\ref{eq:tdse}), as we explain later.
With Eq.~(\ref{eq:C}), the leading-order corrections for the field-modulated quantities are computed as 
\begin{align}
\epsilon^\prime_{n}&=\epsilon_{n}-\sum_{m\neq n}\frac{|e\bm{E}\cdot\bm{\xi}_{nm}|^{2}}{\Delta_{nm}},\label{eq:epsprime}\\
\bm{a}^\prime_{n}&=\bm{a}_{n}+e\sum_{m\neq n}\frac{\bm{E}\cdot\bm{\xi}_{nm}\bm{\xi}_{mn}+\bm{\xi}_{nm}\bm{E}\cdot\bm{\xi}_{mn}}{\Delta_{nm}},\label{eq:aprime}\\
\bm{\xi}_{nm}^{\prime} & =\bm{\xi}_{nm}-ie\hat{\bm{D}}_{nm}\left(\frac{\bm{E}\cdot\bm{\xi}_{nm}}{\Delta_{nm}}\right) \nonumber \\
  &+e\sum_{l\neq n,m}\left(\frac{\bm{E}\cdot\bm{\xi}_{nl}}{\Delta_{nl}}\bm{\xi}_{lm}-\bm{\xi}_{nl}\frac{\bm{E}\cdot\bm{\xi}_{lm}}{\Delta_{lm}}\right),
\end{align}
The tensor in the first-order correction to the Berry connection is sometimes
called the Berry-connection polarizability.

We note that the field-modulated eigenenergy Eq.~(\ref{eq:epsprime}) reads $\epsilon^\prime_{n}=\dot{\gamma}_{n\bm{k}_{0}}^{(0)}-\dot{\gamma}_{n\bm{k}_{0}}^{(2)}$ with a sign flip in the correction term, while sometimes the quantity corresponding
to $\dot{\gamma}^{(0)}+\dot{\gamma}^{(2)}$ is interpreted as the
gauge-invariant part of the eigenenergy in the length-gauge formalism (given by $\dot{\gamma}^\prime$). From the viewpoint of adiabatic
time evolution, the proper form can be deduced from the fact that the phase
factor $\gamma^\prime$ consists of the dynamical and Berry phase contributions,
and $\gamma^{(2)}$ should also be composed of corrections to both
contributions in general. The appearance of the Berry phase factor
in the adiabatic wave function $|\psi_{n\bm{k}_{0}}^{\text{ad}}(t)\rangle$
stems from the fact that the overall phase of $|\psi_{n\bm{k}_{0}}^{\text{ad}}(t)\rangle$
includes that of $|u_{n\bm{k}(t)}\rangle$ in an implicit way [See
Eq.~(\ref{eq:ad})]. Because the overall phase of $|u_{n\bm{k}(t)}\rangle$
is gauge-dependent and sometimes not smooth, one needs a counter factor
to realize a smooth time evolution generated by Eq.~(\ref{eq:tdse}).
Indeed, due to the Berry phase factor, we have
\begin{align}
\frac{d}{dt}\left(|u_{n\bm{k}(t)}\rangle e^{-i\gamma_{n\bm{k}_{0}}^{(1)}(t)}\right) & =e^{-i\gamma_{n\bm{k}_{0}}^{(1)}(t)}(-e\bm{E}\cdot\nabla_{\bm{k}}-i\dot{\gamma}_{n\bm{k}_{0}}^{(1)})|u_{n\bm{k}(t)}\rangle\nonumber \\
 & =ie\sum_{m\neq n}|u_{m\bm{k}(t)}\rangle e^{-i\gamma_{n\bm{k}_{0}}^{(1)}(t)}\bm{E}\cdot\bm{\xi}_{mn}(\bm{k}(t))\label{eq:xi}
\end{align}
for the unperturbed adiabatic evolution, which indicates that the temporal change is purely offdiagonal and
includes no phase shift. Thus, the total evolution of the overall phase
is described solely by the dynamical phase factor, and only this
phase may affect physical properties. Similarly, the field-modulated adiabatic wave function $|\psi^\prime_{n\bm{k}_{0}}(t)\rangle$ also depends implicitly on the overall phase of the basis vector
$|u^\prime_{n\bm{k}(t)}\rangle$. The rate of the overall phase shift 
is given by $\langle\psi^\prime_{n\bm{k}_{0}}(t)|i\partial_{t}|\psi^\prime_{n\bm{k}_{0}}(t)\rangle=\dot{\gamma}^\prime_{n\bm{k}_{0}}(t)-e\bm{E}\cdot\bm{a}^\prime_{n}(\bm{k}(t))$
from Eqs.~(\ref{eq:psi-2}) and (\ref{eq:aprime-2}), which coincides with $\langle\psi^\prime_{n\bm{k}_{0}}(t)|i\partial_{t}|\psi^\prime_{n\bm{k}_{0}}(t)\rangle=\langle u^\prime_{n\bm{k}(t)}|H(\bm{k}(t))|u^\prime_{n\bm{k}(t)}\rangle$ from Eq.~(\ref{eq:tdse}). Because we
can easily show $e\bm{E}\cdot\bm{a}^\prime_{n}=\dot{\gamma}_{n\bm{k}_{0}}^{(1)}+2\dot{\gamma}_{n\bm{k}_{0}}^{(2)}$
using Eq.~(\ref{eq:aprime}), we finally have $\epsilon_n^\prime=\dot{\gamma}_{n\bm{k}_{0}}^{(0)}-\dot{\gamma}_{n\bm{k}_{0}}^{(2)}$ with the opposite sign as physical energy. A similar suggestion has also been proposed in Ref.~\cite{Tang2026} based on the polarization work of the electric field.

While, as we see later, the second-order eigenenergy correction $\epsilon^\prime$ does
not contribute to the second-order electric current response~\cite{Gao2014,Xiao2025,Qiang2026,Tang2026}, 
the misinterpretation of the eigenenergy might lead to incorrect results, 
especially in higher-order responses [e.g., with the $O(E)$ drift of the distribution].

\section{Nonlinear current response}\label{sec:current}

In order to calculate the DC current response of the system, we need to determine the reduced density matrix of the system in the presence of the applied electric field. 
While phenomenological expressions based on Boltzmann transport or relaxation-time approximations are employed in  length-gauge studies~\cite{Watanabe2020,Kaplan2024,Qiang2026,Tang2026}, here we instead adopt the nonequilibrium Green function approach and determine the steady state in a fully quantum framework, as in recent velocity-gauge studies reporting the longitudinal nonlinear current~\cite{Ulrich2026,Guo2026,Anan2026}.

Here we consider the system coupled to a fermionic reservoir~\cite{Buttiker1986,Jauho1994},
represented as
\begin{align}
\mathcal{H}_{\text{tot}} & =\sum_{\bm{k}_{0}\alpha\beta}\langle\alpha|H(\bm{k}_{0}+e\bm{A}(t))|\beta\rangle\hat{c}_{\bm{k}_{0}\alpha}^{\dagger}(t)\hat{c}_{\bm{k}_{0}\beta}(t)\nonumber \\
 & +\sum_{\bm{k}_{0}\alpha p}(V_{p}\hat{b}_{\bm{k}_{0}\alpha p}^{\dagger}(t)\hat{c}_{\bm{k}_{0}\alpha}(t)+V_{p}^{\ast}\hat{c}_{\bm{k}_{0}\alpha}^{\dagger}(t)\hat{b}_{\bm{k}_{0}\alpha p}(t))\nonumber \\
 & +\sum_{\bm{k}_{0}\alpha p}\omega_{p}\hat{b}_{\bm{k}_{0}\alpha p}^{\dagger}(t)\hat{b}_{\bm{k}_{0}\alpha p}(t).\label{eq:htot}
\end{align}
Here $\hat{c},\hat{b}$ are the fermionic field operators for the
system and reservoir, respectively, with the Heisenberg picture. $|\beta\rangle$
is the one-particle basis for the internal degree of freedom, which
does not depend on time $t$ nor momentum $\bm{k}_{0}$. The spectral
density of the reservoir is chosen as
\begin{equation}
\pi\sum_{p}|V_{p}|^{2}\delta(\omega-\omega_{p})=\Gamma=\text{const}.
\end{equation}

The nonequilibrium Green functions, defined as $\langle\alpha|G_{\bm{k}_{0}}^{R}(t,t^{\prime})|\beta\rangle=-i\langle\{\hat{c}_{\bm{k}_{0}\alpha}(t),\hat{c}_{\bm{k}_{0}\beta}^{\dagger}(t^{\prime})\}\rangle\Theta(t-t^{\prime})$,
$G_{\bm{k}_{0}}^{A}(t,t^{\prime})=[G_{\bm{k}_{0}}^{R}(t^{\prime},t)]^{\dagger}$, and $\langle\alpha|G_{\bm{k}_{0}}^{<}(t,t^{\prime})|\beta\rangle=i\langle\hat{c}_{\bm{k}_{0}\beta}^{\dagger}(t^{\prime})\hat{c}_{\bm{k}_{0}\alpha}(t)\rangle$,
for the present system are given by
\begin{align}
G_{\bm{k}_{0}}^{R}(t,t^{\prime}) & =-iU(t,t^{\prime})e^{-\Gamma(t-t^{\prime})}\Theta(t-t^{\prime}),\\
G_{\bm{k}_{0}}^{<}(t,t^{\prime}) & =\int d\tau\int d\tau^{\prime}G_{\bm{k}_{0}}^{R}(t,\tau)\Sigma^{<}(\tau,\tau^{\prime})G_{\bm{k}_{0}}^{A}(\tau^{\prime},t^{\prime}),\\
\Sigma^{<}(t,t^{\prime}) & =2i\Gamma\int\frac{d\omega}{2\pi}f_D(\omega)e^{-i\omega(t-t^{\prime})}
\end{align}
with $f_D$ being the Fermi-Dirac distribution function.
Here the one-particle time-evolution operator $U(t,t^{\prime})$ for
the isolated system can be written by the field-modulated adiabatic wave function $|\psi^\prime_{n\bm{k}_{0}}(t)\rangle$
as
\begin{align}
U(t,t^{\prime}) & =\sum_{n}|\psi^\prime_{n\bm{k}_{0}}(t)\rangle\langle\psi^\prime_{n\bm{k}_{0}}(t^{\prime})|.
\end{align}

We introduce the field operator in the (field-modulated) adiabatic basis $\hat{\psi}^\prime_{n\bm{k}_{0}}(t)$
by~\cite{Kitamura2020} 
\begin{align}
\hat{\psi}^\prime_{n\bm{k}_{0}}(t) & =\sum_{\alpha}\langle\psi^\prime_{n\bm{k}_{0}}(t)|\alpha\rangle\hat{c}_{\bm{k}_{0}\alpha}(t),\\
\hat{c}_{\bm{k}_{0}\alpha}(t) & =\sum_{n}\langle\alpha|\psi^\prime_{n\bm{k}_{0}}(t)\rangle\hat{\psi}^\prime_{n\bm{k}_{0}}(t).
\end{align}
Then the current expectation value in this basis is given by
\begin{align}
\bm{J} & =-e\int\frac{d^{3}\bm{k}_{0}}{(2\pi)^{3}}\sum_{\alpha\beta}\langle\beta|\nabla_{\bm{k}}H(\bm{k}(t))|\alpha\rangle\langle\hat{c}_{\bm{k}_{0}\beta}^{\dagger}(t)\hat{c}_{\bm{k}_{0}\alpha}(t)\rangle\nonumber \\
 & =-e\int\frac{d^{3}\bm{k}_{0}}{(2\pi)^{3}}\sum_{nm}\bm{V}^\prime_{nm}(t)\langle\hat{\psi}_{n\bm{k}_{0}}^{\prime\dagger}(t)\hat{\psi}^\prime_{m\bm{k}_{0}}(t)\rangle,
\end{align}
where the velocity matrix element $\bm{V}^\prime_{nm}(t)$ has already been
derived in Eq.~(\ref{eq:V}). The remaining quantity to identify
is the lesser Green function in the adiabatic basis 
\begin{align}
[{G}_{\bm{k}_{0}}^{<}(t,t^{\prime})]_{nm} & =i\langle\hat{\psi}_{m\bm{k}_{0}}^{\prime\dagger}(t^{\prime})\hat{\psi}^\prime_{n\bm{k}_{0}}(t)\rangle\nonumber \\
 & =\langle\psi^\prime_{n\bm{k}_{0}}(t)|G_{\bm{k}_{0}}^{<}(t,t^{\prime})|\psi^\prime_{m\bm{k}_{0}}(t^{\prime})\rangle
\end{align}
for $t=t^{\prime}$. The retarded Green function in the adiabatic
basis takes the form 
\begin{align}
\langle\psi^\prime_{n\bm{k}_{0}}(t)|G_{\bm{k}_{0}}^{R}(t,t^{\prime})|\psi^\prime_{m\bm{k}_{0}}(t^{\prime})\rangle & =-i\delta_{nm}e^{-\Gamma(t-t^{\prime})}\Theta(t-t^{\prime}),
\end{align}
with which we can write the lesser component as
\begin{align}
[{G}_{\bm{k}_{0}}^{<}(t,t^{\prime})]_{nm} & =\int_{0}^{\infty}d\tau\int_{0}^{\infty}d\tau^{\prime}[{\Sigma}_{\bm{k}_{0}}^{<}(t-\tau,t^{\prime}-\tau^{\prime})]_{nm}e^{-\Gamma(\tau+\tau^{\prime})},\label{eq:lesser}\\{}
[{\Sigma}_{\bm{k}_{0}}^{<}(t,t^{\prime})]_{nm} & =2i\Gamma\int\frac{d\omega}{2\pi}f_D(\omega)\langle\psi^\prime_{n\bm{k}_{0}}(t)|\psi^\prime_{m\bm{k}_{0}}(t^{\prime})\rangle e^{-i\omega(t-t^{\prime})}.\label{eq:sigma-lesser}
\end{align}

The time convolution here can be carried out by means of term-by-term integration, after expanding the wave function $|\psi^\prime_{m\bm{k}_{0}}(t-\tau)\rangle$
into a power series with respect to $\tau$. As we detail in Appendix~\ref{sec:appendix-overlap},
we can compute the overlap of the adiabatic wave function up to $E^2$
as $\langle\psi^\prime_{m\bm{k}_{0}}(t-\tau)|\psi^\prime_{m\bm{k}_{0}}(t-\tau^{\prime})\rangle=S_{m}(\bm{k}(t),\tau,\tau^{\prime})e^{-i\epsilon_{m}(\bm{k}(t))(\tau-\tau^{\prime})}$
with 
\begin{align}
 & S_{m}(\bm{k},\tau,\tau^{\prime})\nonumber \\
 & =1+i\sum_{n\neq m}\frac{|e\bm{E}\cdot\bm{\xi}_{mn}|^{2}}{\Delta_{mn}}(\tau-\tau^{\prime})-\frac{ie}{2}\bm{E}\cdot\nabla_{\bm{k}}\epsilon_{m}(\tau^{2}-\tau^{\prime2})\nonumber \\
 & -\frac{e^{2}}{2}\bm{E}\cdot\bm{g}_{m}\cdot\bm{E}(\tau-\tau^{\prime})^{2}-\frac{ie^{2}}{6}\left(\bm{E}\cdot\nabla_{\bm{k}}\right)^{2}\epsilon_{m}(\tau^{3}-\tau^{\prime3})\nonumber \\
 & -\frac{e^{2}}{8}\left(\bm{E}\cdot\nabla_{\bm{k}}{\epsilon_{m}}\right)^{2}(\tau^{2}-\tau^{\prime2})^{2}+O(E^{3})\label{eq:S}
\end{align}
for the diagonal component. The offdiagonal component of $G^<$ is shown to
be $O(\Gamma)$, see Appendix~\ref{sec:appendix-overlap}.
We find a contribution from the quantum metric $g_{m}^{ij}(\bm{k})$, which is essentially derived from the overlap of the
instantaneous eigenstates
\begin{multline}
\langle u_{n,\bm{k}(t-\tau)}|u_{n,\bm{k}(t-\tau^{\prime})}\rangle e^{i\gamma_{n\bm{k}_{0}}^{(1)}(t-\tau)-i\gamma_{n\bm{k}_{0}}^{(1)}(t-\tau^{\prime})}\\
=1-\frac{e^{2}}{2}\bm{E}\cdot\bm{g}_{n}\cdot\bm{E}(\tau-\tau^{\prime})^{2}+O(E^3),\label{eq:metric-origin}
\end{multline}
which follows from Eq.~(\ref{eq:xi}), while the remaining terms come from 
the expansion of the dynamical phase factor.

Using the overlap matrix element obtained above, we can perform the
frequency integral by the following formula
\begin{align}
 & 2\Gamma\int_{0}^{\infty}d\tau\int_{0}^{\infty}d\tau^{\prime}\int\frac{d\omega}{2\pi}f_D(\omega)(\tau+\tau^{\prime})^{n}(\tau-\tau^{\prime})^{m}\nonumber \\
 & \times e^{i(\omega-x)(\tau-\tau^{\prime})-\Gamma(\tau+\tau^{\prime})}=\frac{n!}{\Gamma^{n}}i^{m}\frac{\partial^{m}f_D(x)}{\partial x^{m}}+O\left(\frac{\Gamma}{k_BT}\right).\label{eq:freq}
\end{align}
We finally arrive at
\begin{align}
 & \langle\hat{\psi}_{m\bm{k}_{0}}^{\prime\dagger}(t)\hat{\psi}^\prime_{m\bm{k}_{0}}(t)\rangle\nonumber \\
 & =f_D(\epsilon_{m})+\frac{e}{2\Gamma}\bm{E}\cdot\nabla_{\bm{k}}f_D(\epsilon_{m})+\frac{e^{2}}{4\Gamma^{2}}\left(\bm{E}\cdot\nabla_{\bm{k}}\right)^{2}f_D(\epsilon_{m})\nonumber \\
 & -\sum_{n\neq m}\frac{|e\bm{E}\cdot\bm{\xi}_{mn}|^{2}}{\Delta_{mn}}f_D^{\prime}(\epsilon_{m})-\frac{e^{2}}{24}\left(\bm{E}\cdot\nabla_{\bm{k}}\right)^{2}\epsilon_{m}f_D^{\prime\prime\prime}(\epsilon_{m})\nonumber \\
 & +\frac{e^{2}}{2}\bm{E}\cdot\bm{g}_{m}\cdot\bm{E}f_D^{\prime\prime}(\epsilon_{m}).\label{eq:f}
\end{align}
Here the first four terms reproduce the semiclassical expression with the Boltzmann equation, while the last two terms appear as quantum corrections.
Now, we can complete the calculation of the nonlinear current. This yields 
\begin{align}
\bm{J} & =-e\sum_{n}\int\frac{d^{3}\bm{k}_{0}}{(2\pi)^{3}}\bm{V}_{n}^{\prime\text{diag}}(\bm{k}(t))\langle\hat{\psi}_{n\bm{k}_{0}}^{\prime\dagger}(t)\hat{\psi}^\prime_{n\bm{k}_{0}}(t)\rangle\nonumber \\
 & =-e\sum_{n}\int_{\bm{k}}\nabla_{\bm{k}}\epsilon_{n}\left[f_{n}+\frac{e}{2\Gamma}\bm{E}\cdot\nabla_{\bm{k}}f_{n}+\frac{e^{2}}{4\Gamma^{2}}(\bm{E}\cdot\nabla_{\bm{k}})^{2}f_{n}\right]\nonumber \\
 & -e^{2}\sum_{n}\int_{\bm{k}}\bm{E}\times\left[\nabla_{\bm{k}}\times\bm{a}^\prime_{n}-\frac{e}{2\Gamma}(\bm{E}\cdot\nabla_{\bm{k}})(\nabla_{\bm{k}}\times\bm{a}_n)\right]f_{n}\nonumber \\
 & -e^{3}\sum_{n}\int_{\bm{k}}\nabla_{\bm{k}}\epsilon_{n}\left[\frac{1}{2}\bm{E}\cdot\bm{g}_{n}\cdot\bm{E}f_{n}^{\prime\prime}-\frac{1}{24}(\bm{E}\cdot\nabla_{\bm{k}})^{2}\epsilon_{n}f_{n}^{\prime\prime\prime}\right]\nonumber \\
 & +e^{3}\sum_{n}\sum_{m\neq n}\int_{\bm{k}}\nabla_{\bm{k}}\left(\frac{|\bm{E}\cdot\bm{\xi}_{nm}|^{2}}{\Delta_{nm}}f_{n}\right)
\end{align}
with the shorthand $f_{n}=f_D(\epsilon_{n})$ and $\int_{\bm{k}}=\int d^{3}\bm{k}/(2\pi)^{3}$.
Here, in the last line, the longitudinal current contribution from the field
modulation of the group velocity is canceled out by the correction
to the distribution function, in the form of $\int_{\bm{k}}\nabla_{\bm{k}}(\dots)=0$.
Since these contributions represent the second-order part of the electric current of the equilibrium state (with modulated eigenenergy) $\int_{\bm{k}}\nabla_{\bm{k}}\epsilon^\prime_nf_D(\epsilon^\prime_n)$, they always vanish regardless of the detailed form. This is consistent with recent semiclassical studies~\cite{Xiao2025,Qiang2026,Tang2026}.
However, due to the quantum-metric correction to the distribution
function in the last line of Eq.~(\ref{eq:f}), we obtain the longitudinal nonreciprocal current 
in terms of the quantum-metric dipole as 
\begin{multline}
-\frac{e^{3}}{2}\sum_{n}\int\frac{d^{3}\bm{k}}{(2\pi)^{3}}\nabla_{\bm{k}}\epsilon_{n}\bm{E}\cdot\bm{g}_{n}\cdot\bm{E}f_D^{\prime\prime}(\epsilon_{n})\\
=\frac{e^{3}}{2}\sum_{n}\int\frac{d^{3}\bm{k}}{(2\pi)^{3}}\nabla_{\bm{k}}\left(\bm{E}\cdot\bm{g}_{n}\cdot\bm{E}\right)f_D^{\prime}(\epsilon_{n})
\end{multline}
with the use of integration by parts. The entire expression for the nonlinear current obtained
here coincides exactly (including coefficients) with that of Ref.~\cite{Ulrich2026} at zero temperature, which also adopts the full-quantum approach with the velocity gauge. 
Also, we can rewrite the third-derivative term due to the quantum correction as
\begin{align}
 & \frac{e^{3}}{24}\sum_{n}\int\frac{d^{3}\bm{k}}{(2\pi)^{3}}\nabla_{\bm{k}}\epsilon_{n}(\bm{E}\cdot\nabla_{\bm{k}})^{2}\epsilon_{n}f_D^{\prime\prime\prime}(\epsilon_{n})\nonumber \\
 & =-\frac{e^{3}}{24}\sum_{n}\int\frac{d^{3}\bm{k}}{(2\pi)^{3}}\nabla_{\bm{k}}\epsilon_{n}(\bm{E}\cdot\nabla_{\bm{k}})^{2}f_{D}^{\prime\prime}(\epsilon_{n})\nonumber \\
 & =-\frac{e^{3}}{48}\sum_{n}\int\frac{d^{3}\bm{k}}{(2\pi)^{3}}\nabla_{\bm{k}}\epsilon_{n}(\bm{E}\cdot\nabla_{\bm{k}}\epsilon_{n})^{2}f_{D}^{\prime\prime\prime\prime}(\epsilon_{n}),
\end{align}
This establishes the equivalence of the present expression with that in Ref.~\cite{Guo2026} as well as Ref.~\cite{Ulrich2026}. This term can be interpreted as the $O(\Gamma^2)$ correction to the $O(\Gamma^{-2})$ nonlinear Drude term due to the smearing of the Fermi distribution function. The same term was also obtained in Ref.~\cite{Matsyshyn2023} for a single Floquet band without wave-function geometry.

We note that the field-modulated distribution function Eq.~(\ref{eq:f}) also changes the electron density. This change must be compensated by a shift in the chemical potential $\mu$ when the dissipator conserves particle number, unlike the fermionic reservoir considered here. However, since the shift in $\mu$ is of second order in $E$, the expression for the second-order nonlinear current obtained here remains valid in such cases.

\section{Origin of the quantum-metric correction}\label{sec:origin}

\begin{figure}
    \centering
    \includegraphics[width=\linewidth]{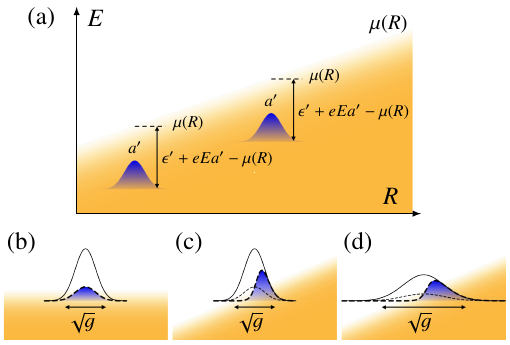}
    \caption{Electron occupation under the tilted chemical potential $\mu(\bm{R})$. 
    (a) Schematic illustration of the gauge-dependent eigenenergy in the length gauge. In the presence of an applied electric field $\bm{E}$, the eigenenergy is shifted according to the center-of-mass position through the Berry connection $a^\prime$. This gauge dependence does not affect physical observables if the chemical potential is tilted accordingly. 
    (b) Local occupation of a Bloch state in the absence of electric fields. 
    The energy of the Bloch state is chosen slightly above the Fermi level. 
    The blue shaded region represents the local occupation, given by the probability density (solid line) multiplied by the Fermi distribution function $f$, whose area gives the total occupation number.
    (c) Local occupation under the tilted chemical potential. The chemical potential at the center of the wave packet is the same as in panel (b). Since $f$ becomes position dependent, one tail of the wave packet becomes nearly fully occupied while the opposite tail remains nearly empty, resulting in a slight increase in the total occupation.    
    (d) Local occupation for a wave packet with a larger spread $\sqrt{g}$. Since $f$ changes from zero to unity over a length scale of $\sim k_BT/(eE)$, the imbalance between the two tails becomes more pronounced for a wider wave packet, leading to a larger change in the total occupation.
    }
    \label{fig:metric}
\end{figure}

In this section, let us focus on the quantum-metric
correction to the distribution function Eq.~(\ref{eq:f}) and discuss its origin.
The quantum metric correction is derived from Eq.~(\ref{eq:metric-origin}) in the velocity-gauge picture, which dictates the deformation of the instantaneous eigenstate between scattering events owing to the drift dynamics.
To the best of our knowledge, this correction has been overlooked so far in the length-gauge formalism.
One possible reason behind this may lie in the nontrivial action of the position operator on the Bloch states, which renders the microscopic description of the relaxation process unexpectedly intricate.
In line with this, the appearance of the quantum metric representing the positional fluctuation implies the relevance of the position operator to the relaxation process.

In the length-gauge picture, the ``eigenenergy" of the perturbed wave function $\epsilon^{\prime(\text{L})}$ is obtained in the same form as $\dot{\gamma}^\prime$, 
\begin{align}
\epsilon_m^{\prime(\text{L})}(\bm{k})&=\epsilon^\prime_m(\bm{k})+e\bm{E}\cdot\bm{a}^\prime_m(\bm{k})\nonumber\\
&=\epsilon_m(\bm{k})+e\bm{E}\cdot\bm{a}_m(\bm{k})+\sum_{n\neq m}\frac{|e\bm{E}\cdot\bm{\xi}_{mn}(\bm{k})|^2}{\Delta_{mn}(\bm{k})},\label{eq:eps-len}
\end{align}
which is explicitly gauge-dependent. The gauge dependence physically comes from the fact that the energy depends on the center-of-mass position in the presence of the voltage bias, while the position of the plane wave is essentially indefinite. On the other hand, the applied voltage bias should also shift the chemical potential at each site by the bias potential, as $\mu(\bm{R})=\mu+\delta\mu(\bm{R})=\mu+e\bm{E}\cdot\bm{R}$, in order to retain the invariance under the electric translation (i.e., translation along with energy shift). This shift guarantees the energy measured from the chemical potential to be gauge-invariant, which is schematically drawn in Fig.~\ref{fig:metric}(a).

In light of this fact, the quantum-metric correction can be interpreted as the bias-induced positional fluctuation of the occupation. 
As we detail in Appendix~\ref{sec:appendix-justification}, one can derive the quantum Boltzmann equation for the present system in the length-gauge formalism, which has the form of the relaxation-time equation
\begin{equation}
-e\bm{E}\cdot\nabla_{\bm{k}}f_{n\bm{k}}=-2\Gamma(f_{n\bm{k}}-f_{n\bm{k}}^{\text{qe}})\label{eq:rta}
\end{equation}
with the quasiequilibrium distribution function $f_{n\bm{k}}^{\text{qe}}$ incorporating quantum corrections.
The quasiequilibrium distribution function under the tilted chemical potential $\delta\mu=e\bm{E}\cdot\hat{\bm{R}}$ is given approximately by
\begin{align}
f_{n\bm{k}}^{\text{qe}} & =\overline{\left\langle f_{D}(\epsilon_{n}^{\prime(\text{L})}(\bm{k})+e\bm{E}\cdot\bm{R}_0-\delta\mu)\right\rangle }_{n\bm{k};\bm{R}_{0}}\nonumber\\
&=\overline{\left\langle f_{D}(\epsilon_{n}^{\prime}(\bm{k})-e\bm{E}\cdot(\hat{\bm{R}}-\langle\hat{\bm{R}}\rangle_{n\bm{k};\bm{R}_{0}}))\right\rangle }_{n\bm{k};\bm{R}_{0}}\nonumber\\
 & \sim f_{D}(\epsilon_{n}^{\prime})+\frac{1}{2}f_{D}^{\prime\prime}(\epsilon_{n}^{\prime})\overline{\left\langle [e\bm{E}\cdot(\hat{\bm{R}}-\langle\hat{\bm{R}}\rangle_{n\bm{k};\bm{R}_{0}})]^{2}\right\rangle }_{n\bm{k};\bm{R}_{0}}.\label{eq:intuitive}
\end{align}
Here we introduce the site-resolved expectation value $\langle\hat{O}\rangle_{n\bm{k};\bm{R}_{0}}$, 
in which the Bloch wave $|u^\prime_{n\bm{k}}\rangle$ is decomposed into Wannier orbitals at site $\bm{R}_0$. For any function of the position operator $F(\hat{\bm{R}})$, it is evaluated as
\begin{align}
\langle F(\hat{\bm{R}})\rangle_{n\bm{k};\bm{R}_{0}} & =\text{Re}\langle u_{n\bm{k}}^{\prime}|F(i\nabla_{\bm{k}^{\prime}}+\bm{R}_{0})|u_{n\bm{k}^{\prime}}^{\prime}\rangle|_{\bm{k}^{\prime}=\bm{k}},\label{eq:site-resolved-R}
\end{align}
while $\overline{F(\bm{R}_{0})}=N^{-1}\sum_{\bm{R}_{0}}F(\bm{R}_{0})$ represents the spatial average. See Appendix~\ref{sec:appendix-justification} for more details.
The gauge-dependent part $\langle \hat{\bm{R}}\rangle_{n\bm{k};\bm{R}_{0}} =\bm{a}_{n}^\prime(\bm{k}) +\bm{R}_{0}$ cancels with the chemical potential shift at the first order, while the second-order fluctuation in the above sense is given by the quantum metric $\bm{g}$ as
\begin{align}
\overline{\left\langle [e\bm{E}\cdot(\hat{\bm{R}}-\langle\hat{\bm{R}}\rangle_{n\bm{k};\bm{R}_{0}})]^{2}\right\rangle }_{n\bm{k};\bm{R}_{0}} & =e^{2}\bm{E}\cdot\bm{g}_{n}\cdot\bm{E}.
\end{align}
Restoring the third-derivative term omitted in the approximate treatment above, the quasiequilibrium distribution reads
\begin{align}
f_{n\bm{k}}^{\text{qe}} & =f_{D}(\epsilon_{n}^{\prime}(\bm{k}))-\frac{e^{2}}{24}(\bm{E}\cdot\nabla_{\bm{k}})^{2}\epsilon_{n}^{\prime}(\bm{k})f_{D}^{\prime\prime\prime}(\epsilon_{n}^{\prime}(\bm{k}))\nonumber\\
&+\frac{e^{2}}{2}\bm{E}\cdot\bm{g}_{n}(\bm{k})\cdot\bm{E}f_{D}^{\prime\prime}(\epsilon_{n}^{\prime}(\bm{k}))+O(E^3).\label{eq:qe-2}
\end{align}
Combined with the relaxation-time equation Eq.~(\ref{eq:rta}), this quasiequilibrium distribution reproduces Eq.~(\ref{eq:f}).
While we focus on the second-order contribution here, the relaxation-time description can be generalized to arbitrary order in $E$, as shown in Appendix~\ref{sec:appendix-higher}.

As clarified here, the quantum-metric correction to the nonequilibrium distribution function is traced back to Eq.~(\ref{eq:intuitive}), which shows that the occupation of the Bloch state is given by the spatial integral of the local contributions. This expression admits an intuitive picture for the metric correction, as sketched in Figs.~\ref{fig:metric}(b)--(d).
Here we consider a Bloch wave packet (drawn by solid lines) with an energy slightly above the Fermi level. The blue shaded region represents the local occupation of the Bloch state as a function of position, given by the probability density multiplied by the Fermi distribution function $f_D$. Accordingly, the area of the blue region represents the total occupation number. While $f_D$ is spatially uniform in panel (b), the tilted chemical potential due to the applied field replaces $f_D(\epsilon^\prime)$ with $f_D(\epsilon^\prime-e\bm{E}\cdot\Delta\bm{R})$ in panels (c) and (d). Since the wave packet is located near the Fermi level, this leads to a drastic change in the local occupation. 
Namely, around the center of the wave packet, local Fermi distribution $f_D$ changes from zero to unity over a length scale of $\sim k_BT/(eE)$, leaving one tail nearly fully occupied while the opposite tail remains nearly empty. This effect becomes relevant when the wave packet has a large spread $\sim\sqrt{g}$, as exemplified in panel (d).

\section{Comparison to other formalisms}\label{sec:comparison}

\begin{table*}[t]
\centering
\begin{tabular*}{\textwidth}{@{\extracolsep{\fill}} llccccccc}
\toprule
Treatment & 
Dissipation & 
$\rho_{\bm{k}}$ & $K_1$ & $K_2$ & 
$\epsilon$ correction to $v$ & 
$\epsilon$ correction to $f$ & 
$g$ correction to $f$ & 
Longitudinal current \\
\colrule
Semiclassical & Conventional RTA & Eq.~(\ref{eq:rho-rta}) & $1$ & $1$ & $\checkmark$ & $-$ & $-$ & Allowed \\ 
Quantum & Conventional RTA & Eq.~(\ref{eq:rho-qrta}) & $1$ & $(0)$ & $\checkmark$ & $-$ & $-$ & Allowed \\ 
Quantum & Adiabatic ramp & Eq.~(\ref{eq:rho-iso}) & $2$ & $3$ & $\checkmark$ & $-$ & $-$ & Allowed \\ 
Semiclassical & Modified RTA & Eq.~(\ref{eq:rho-mrta}) & $1$ & $0$ & $\checkmark$ & $\checkmark$ & $-$ & Absent \\ 
Quantum & Constant self-energy & Eq.~(\ref{eq:rho-metric}) & $1$ & $0$ & $\checkmark$ & $\checkmark$ & $\checkmark$ & Allowed \\ 
\botrule
\end{tabular*}
\caption{
Comparison of the relaxation-time independent nonlinear conductivity for various formalisms. 
Differences among the formalisms can be classified by how the dissipation is incorporated, and summarized in the form of the density matrix $\rho_{\bm{k}}$. 
The values of $K_1,K_2$ for the contribution of the Berry-connection polarizability defined in Eq.~(\ref{eq:tensor}) are displayed for each approach.
In this notation, $K_1$ and $K_2$ are related to the Ohmic $K_O$ and Hall $K_H$ components through $K_O=K_2/3$ and $K_H=K_1-K_2/3$.
Here $(0)$ in the second row means that a residual term of a similar form is present.
We indicate whether corrections derived from the eigenenergy $\epsilon$ and quantum metric $g$ to the group velocity $v$ and distribution function $f$ are included. Except for the Modified RTA (relaxation-time approximation), the longitudinal nonlinear current is allowed, because of $K_2\neq0$ or other mechanisms.
The density matrices listed here arise in the frameworks based on the Boltzmann equation and the Liouville equation, as well as the Green function approach with Keldysh or Matsubara formalism.
}
\label{tab:coeff}
\end{table*}

While the longitudinal current of quantum-metric origin is identified
as a consequence of the shifted quasiequilibrium, the existence of the
longitudinal current originating from the Berry-connection polarizability (BCP) 
is also under debate~\cite{Das2023,Wang2024,Kaplan2024,Xiao2025,Qiang2026,Tang2026}. Specifically, the relaxation-time independent
contribution to the nonlinear conductivity of the form 
\begin{equation}
\sigma_{n}^{i;j,k}=[K_1(\partial^{j}G_{n}^{ik}+\partial^{k}G_{n}^{ij})+(K_2-2K_1)\partial^{i}G_{n}^{jk}]f_{n} 
\label{eq:tensor}
\end{equation}
resolved in band $n$ and momentum $\bm{k}$, in units of $e^{3}/\hbar$,
is proposed with various coefficients $K_1,K_2$~\cite{
Watanabe2020,Michishita2022,Oiwa2022,
Das2023,Wang2024,Kaplan2024,Xiao2025,Qiang2026,Tang2026,Ulrich2026,Guo2026}. Here
$G_{n}^{jk}=\sum_{m\neq n}(\xi_{nm}^{j}\xi_{mn}^{k}+\xi_{nm}^{k}\xi_{mn}^{j})/2\Delta_{nm}$
is (a half of) the BCP in the tensor form. 
We subtract $2K_1$ from $K_2$ for later convenience, with which $K_1=1$ corresponds to the contribution of the field-induced Berry curvature $\nabla_{\bm{k}}\times\bm{a}^\prime$. As one can decompose the contribution of $G_n$ into the Ohmic $K_O$ and Hall $K_H$ components by $K_O=K_2/3$ and $K_H=K_1-K_2/3$, $K_2\neq0$ implies the existence of the longitudinal current.
While the origin of the discrepant coefficients has not fully been understood so far 
because of the variety of formulations involving the two gauge choices, 
here we clarify the origin as the difference in the density matrix, which we summarize in Table~\ref{tab:coeff}. 

\subsection{Conventional approach in semiclassics}\label{sec:rta}
As pointed out in recent studies~\cite{Xiao2025,Qiang2026,Tang2026}, the nonzero longitudinal current with
$K_2\neq0$ in several studies is attributed to the lack of eigenenergy correction
to the (quasi)equilibrium distribution of the relaxation-time
approximation (RTA) for the semiclassical Boltzmann equation 
\begin{equation}
-e\bm{E}\cdot\nabla_{\bm{k}}f_{m}=-\frac{f_{m}-f_{m}^{\text{qe}}}{\tau}.
\end{equation}
When the conventional RTA with the unperturbed equilibrium distribution $f^{\text{qe}}_m=f_D(\epsilon_m)$ is adopted for the modulated wave function,
the (reduced) density matrix $\rho_{\bm{k}}$ for momentum $\bm{k}$ reads
\begin{align}
 \rho_{\bm{k}}  & =\sum_m|u^\prime_{m\bm{k}}\rangle\langle u^\prime_{m\bm{k}}| \left[1+e\tau\bm{E}\cdot\nabla_{\bm{k}}+e^2\tau^2(\bm{E}\cdot\nabla_{\bm{k}})^{2}\right]f_D(\epsilon_{m}).\label{eq:rho-rta}
\end{align}
With this, we obtain the longitudinal current with $K_2=1$ from the modulation of the group velocity $\nabla_{\bm{k}} \epsilon^\prime_m=
\nabla_{\bm{k}} \epsilon_m-\nabla_{\bm{k}}G_{n}^{jk}E^jE^k
$, while the transverse current with $K_1=1$ is also obtained with the modulation of the Berry curvature (The first row of Table~\ref{tab:coeff}).
The coefficient $K_2$ can take different values when one adopts an inappropriate form of the group velocity $\nabla_{\bm{k}} \epsilon^\prime_m$. For instance, since the formal ``eigenenergy" in the length-gauge formalism reads $\dot{\gamma}^\prime$ [See Eq.~(\ref{eq:eps-len})] and requires subtraction of the Berry-connection contribution to get a gauge-invariant form, improper subtraction $\dot{\gamma}^\prime_m-e\bm{E}\cdot\bm{a}_m$ rather than $\dot{\gamma}^\prime_m-e\bm{E}\cdot\bm{a}^\prime_m$ results in $K_2=-1$.

\subsection{Modified semiclassical treatment}

As is apparent from Eq.~(\ref{eq:rho-rta}), the conventional RTA for the distribution function leads to unequal treatment of the wave function and energy, since the field-perturbed eigenstates $|u_{m\bm{k}}^\prime\rangle$ are employed to incorporate the anomalous velocity while the eigenenergy remains unperturbed.
If one assumes that the field-modulated eigenstates $|u_{m\bm{k}}^\prime\rangle$ should relax toward the field-modulated equilibrium $f^{\text{qe}}_m=f_D(\epsilon^\prime_{m})$,
the density matrix is modified to~\cite{Xiao2025,Qiang2026,Tang2026}
\begin{align}
 \rho_{\bm{k}}  & =\sum_m|u^\prime_{m\bm{k}}\rangle\langle u^\prime_{m\bm{k}}|\left[1+e\tau\bm{E}\cdot\nabla_{\bm{k}}+e^2\tau^2(\bm{E}\cdot\nabla_{\bm{k}})^{2}\right]f_D(\epsilon^\prime_{m})
 \label{eq:rho-mrta}
\end{align}
with $f_D(\epsilon^\prime_m)=
f_D(\epsilon_m)-G_{m}^{jk}E^jE^kf_D^\prime(\epsilon_m)$.
In this case, the group velocity correction in the conventional case forms a combination 
$\nabla_{\bm{k}}\epsilon_m^\prime f_D(\epsilon_m^\prime)$, which 
can be interpreted as the equilibrium current and thus always vanishes identically ($K_2=0$), regardless of the detailed form of $\epsilon^\prime_m$.
Then the contribution from the BCP should be purely transverse with $K_1=1$ (The fourth row of Table~\ref{tab:coeff}).

While the employment of the quasiequilibrium distribution function $f^{\text{qe}}_m=f_D(\epsilon^\prime_{m})$ here is rather phenomenological, it is consistent with the full-quantum result.
Also, as we see below, the conventional RTA for the quantum Liouville equation includes unphysical contribution~\cite{Terada2024,Terada2025}, suggesting the necessity of field correction.

On the other hand, the above discussion based on semiclassical phenomenology cannot rule out the existence of the longitudinal current response arising from the quantum correction, which is indeed the case in our result. Below, we examine the formalisms based on quantum treatments.

\subsection{Conventional RTA in the quantum framework}\label{sec:qrta}

We first examine the conventional RTA in the quantum framework, 
as a counterpart of the semiclassical treatment discussed in Sec.~\ref{sec:rta}. 
The quantum Liouville equation with the phenomenological relaxation toward the unperturbed equilibrium 
\begin{align}
\rho_{\bm{k}}^{\text{qe}}&=\sum_{m}|u_{m\bm{k}}\rangle\langle u_{m\bm{k}}|f_D(\epsilon_{m}(\bm{k}))
\end{align}
reads
\begin{align}
\dot{\rho}_{\bm{k}_{0}}(t)	&=-i[H(\bm{k}(t)),\rho_{\bm{k}_{0}}(t)]-\frac{\rho_{\bm{k}_{0}}(t)-\rho_{\bm{k}(t)}^{\text{qe}}}{\tau}\label{eq:qrta}
\end{align}
for the velocity gauge. This is equivalent to taking $\rho^{\text{qe}}_{\bm{k}_0}$ as the equilibrium density matrix in the length gauge~\cite{Passos2018}, and the resulting density matrix satisfies the same recursive equation as that in Ref.~\cite{Watanabe2020}. 

Because $|\psi_{m\bm{k}_{0}}^{\prime}(t)\rangle$ satisfies Eq.~(\ref{eq:tdse}), the equation of motion for  $\langle\psi_{m\bm{k}_{0}}^{\prime}(t)|\rho_{\bm{k}_0}(t)|\psi_{n\bm{k}_{0}}^{\prime}(t)\rangle$ can be easily solved. The resultant density matrix for Eq.~(\ref{eq:qrta}), $\rho_{\bm{k}_0}(t)=\rho_{\bm{k}(t)}$, reads 
\begin{multline}
\rho_{\bm{k}}  =\sum_{m}|u_{m\bm{k}}^{\prime}\rangle\langle u_{m\bm{k}}^{\prime}|\left[1+e\tau\bm{E}\cdot\nabla_{\bm{k}}+e^{2}\tau^{2}(\bm{E}\cdot\nabla_{\bm{k}})^{2}\right]\\
\times\langle u_{m\bm{k}}^{\prime}|\rho_{\bm{k}}^{\text{qe}}|u_{m\bm{k}}^{\prime}\rangle+O(\tau^{-1})\label{eq:rho-qrta}
\end{multline}
with
\begin{align}
\langle u_{m\bm{k}}^{\prime}|\rho_{\bm{k}}^{\text{qe}}|u_{m\bm{k}}^{\prime}\rangle &  =f_{D}(\epsilon_{m})-\sum_{n\neq m}\left|\frac{e\bm{E}\cdot\bm{\xi}_{nm}}{\Delta_{nm}}\right|^{2}(f_{D}(\epsilon_{m})-f_{D}(\epsilon_{n})).
\end{align}
The deviation from the equilibrium distribution $f_D(\epsilon_m)$ has a form similar but different from that of $f_D(\epsilon^\prime_m)$.
This yields the nonlinear current with $K_1=1,K_2=0$ (The second row of Table~\ref{tab:coeff}), but with an additional term
\begin{align}
\delta\bm{J} & =e^{3}\sum_{m}\sum_{n\neq m}\int_{\bm{k}}\frac{\nabla_{\bm{k}}|\bm{E}\cdot\bm{\xi}_{nm}|^{2}}{\Delta_{nm}}f_{D}(\epsilon_{n}).
\end{align}
While the contribution to $K_2$ in the semiclassical treatment is partially canceled, the remaining part $\delta\bm{J}$ shown above may contribute to the longitudinal current. 
However, this contribution should be unphysical since this term can be nonzero even for insulating cases.
Actually, along with this nonlinear current, the conventional RTA is reported to exhibit unphysical interband current even in the linear regime~\cite{Terada2024,Terada2025}. This unphysical current can be written in the form
\begin{align}
\bm{J}^{\text{off}}&=-e\sum_{n}\int_{\bm{k}}\frac{\bm{a}_{n}^{\prime}-\bm{a}_{n}}{\tau}f_{D}(\epsilon_{n}),
\end{align}
which corresponds to the field-induced polarization per relaxation time.
Although the unphysical contributions here can be dropped when one focus on the leading-order of $\tau$ (as in Ref.~\cite{Watanabe2020}),
this result clearly indicates that the relaxation should occur toward the shifted equilibrium with the reactive response, as proposed in semiclassics~\cite{Xiao2025,Qiang2026,Tang2026} and quantum master equation approach~\cite{Terada2024,Terada2025}.

\subsection{Adiabatic ramp}\label{sec:ramp}
Meanwhile, several studies conclude $K_1=2,K_2=3$ based on the quantum dynamics of the density matrix~\cite{Das2023,Wang2024}, whose possible quantum origin should also be examined. As we see below, the origin of the discrepancy with other formalisms lies in a prescription sometimes adopted for convenience, namely the mimicking of a dissipation effect (i.e., a finite lifetime of the electron) through an adiabatic ramp of the external field.
While both the adiabatic ramp $\bm{E}(t)=\bm{E}e^{\eta t}$ and the quasiparticle lifetime enter as an imaginary part of the energy denominator and thus seem to play a similar role, the physical state in the ramping scheme is crucially different from the quasiparticle states with finite lifetime, when it is applied for noninteracting electrons with infinite lifetime. 
With the present framework, the density matrix after the adiabatic ramp can be specified by the modulated adiabatic wave function $|\psi^\prime_{m\bm{k}_0}(t)\rangle$ calculated with finite $\eta$, with the initial occupation $f_D(\epsilon_m(\bm{k}(t=-\infty)))$ kept intact due to the absence of dissipation. This form implies the absence of eigenenergy correction to the distribution function.

Let us inspect the explicit form of the density matrix in more detail. Since the vector potential corresponding to the adiabatic ramp is given by $\bm{A}(t)=-\eta^{-1}\bm{E}e^{\eta t}$, 
the density matrix at $t=0$, $\bm{k}=\bm{k}_0+e\bm{A}(t=0)$ reads $\rho_{\bm{k}_0-\eta^{-1}e\bm{E}}=\sum_m[|u^\prime_{m,\bm{k}_0-\eta^{-1}e\bm{E}}\rangle\langle u^\prime_{m,\bm{k}_0-\eta^{-1}e\bm{E}}|]_{\eta\neq0}f_D(\epsilon_m(\bm{k}_0))$. 
This expression indicates that $\rho_{\bm{k}}$ has the momentum shift of $\eta^{-1}e\bm{E}$ in the distribution, 
\begin{align}
& f_D\left(\epsilon_{m}\left(\bm{k}+\frac{e}{\eta}\bm{E}\right)\right)=\left[1+\frac{e}{\eta}\bm{E}\cdot\nabla_{\bm{k}} +\frac{e^{2}}{2\eta^2}\left(\bm{E}\cdot\nabla_{\bm{k}}\right)^{2}\right]f_D(\epsilon_{m}),
\end{align}
which indeed mimics the drift effect at the first order, while the coefficient deviates at the second order.
In addition, finite $\eta$ also induces modulation of the basis function, given by [See Eqs.~(\ref{eq:C}) and (\ref{eq:C-osc1})]
\begin{align}
[\tilde{C}_{nm}]_{\eta\neq0}&=\tilde{C}_{nm}-i\eta e\frac{\bm{E}\cdot\bm{\xi}_{nm}}{\Delta_{nm}^{2}}(1-\delta_{nm})+O(\eta E^2,\eta^2),\label{eq:C-iso}
\end{align}
which comes from the spurious field profile $\dot{\bm{E}}=\eta\bm{E}$ (at $t=0$).
Then the combination of the $1/\eta$ and $\eta$ corrections leads to a finite offdiagonal part in $\eta\to0$, expressed as
\begin{align}
\rho_{\bm{k}} & =\sum_{m}|u^\prime_{m\bm{k}}\rangle\langle u^\prime_{m\bm{k}}|\left[1+\frac{e}{\eta}\bm{E}\cdot\nabla_{\bm{k}}+\frac{e^{2}}{2\eta^{2}}(\bm{E}\cdot\nabla_{\bm{k}})^{2}\right]f_D(\epsilon_{m})\nonumber\\
 & +i\sum_{m}\sum_{n\neq m}|u^\prime_{n\bm{k}}\rangle\langle u^\prime_{m\bm{k}}|e^{2}\frac{\bm{E}\cdot\bm{\xi}_{nm}}{\Delta_{nm}^{2}}\bm{E}\cdot\nabla_{\bm{k}}(f_D(\epsilon_{n})-f_D(\epsilon_{m})).\label{eq:rho-iso}
\end{align}
The resultant offdiagonal current
\begin{equation}
\bm{J}^{\text{off}}=e^3\int_{\bm{k}_{0}}\sum_{m}
 \sum_{n\neq m}\frac{\bm{E}\cdot\bm{\xi}_{nm}\bm{\xi}_{mn}}{\Delta_{nm}}
    \bm{E}\cdot\nabla_{\bm{k}}(f_D(\epsilon_{n})-f_D(\epsilon_{m}))
\end{equation}
leads to the additional contribution characterized by $K_1=1$, $K_2=2$, which adds up to $K_1=2$, $K_2=3$ ($K_H=K_O=1$) with the contribution common to the conventional RTA (The third row of Table~\ref{tab:coeff}).
As the above analysis shows, the additional contribution here comes from the continuously-growing electric field, which is actually absent in the real situation and is an artifact of the adiabatic ramping scheme which was employed to mimic the finite lifetime of the electrons.

\subsection{Full-quantum scheme with constant self-energy}
As discussed above, the conventional approaches in the quantum formulations (Secs.~\ref{sec:qrta} and \ref{sec:ramp}) fail to capture the eigenenergy correction to the distribution function proposed in the semiclassics~\cite{Xiao2025,Qiang2026,Tang2026}. On the other hand, with a full-quantum framework with a constant imaginary self-energy~\cite{Ulrich2026,Anan2026} or coupling to a fermionic reservoir~\cite{Guo2026}, one can incorporate this correction automatically and the contribution of the $K_2$ term to the longitudinal current disappears (with $K_2=0$), which is consistent with the semiclassical treatments with the modified RTA~\cite{Xiao2025,Qiang2026,Tang2026} (The fifth row of Table~\ref{tab:coeff}).
Furthermore, as clarified in the present study, we identify an additional quantum-metric correction to the density matrix given by
\begin{align}
\rho_{\bm{k}}&= \sum_m|u^\prime_{m\bm{k}}\rangle
\langle u^\prime_{m\bm{k}}|\Biggl[
f_D(\epsilon^\prime_m)
+e\tau\bm{E}\cdot\nabla_{\bm{k}}f_D(\epsilon^\prime_m)\nonumber\\
&+e^2\tau^2(\bm{E}\cdot\nabla_{\bm{k}})^2f_D(\epsilon^\prime_m)
-\frac{e^2}{24}(\bm{E}\cdot\nabla_{\bm{k}})^2\epsilon^\prime_m
f_D^{\prime\prime\prime}(\epsilon^\prime_m)\nonumber\\
&
+\frac{e^2}{2}\bm{E}\cdot\bm{g}_m\cdot\bm{E}
f_D^{\prime\prime}(\epsilon^\prime_m)
\Biggr].
\label{eq:rho-metric}
\end{align}
As a result, the longitudinal nonlinear current associated with the quantum-metric dipole is allowed to appear. This has a quantum origin that is not captured in the semiclassical treatment and is not due to an inappropriate treatment of the dissipation mechanism.
Namely, the present mechanism for longitudinal nonlinear current of the quantum-metric origin has a clear physical picture related to the spread of the electron wave packet.

As we have discussed, the density matrix Eq.~(\ref{eq:rho-metric}) can also be reproduced by the quantum Boltzmann equation, which is equivalent to the semiclassical RTA with the quasiequilibrium distribution Eq.~(\ref{eq:qe-2}), $f^{\text{qe}}_m=f_D(\epsilon^\prime_m)-(e^{2}/24)(\bm{E}\cdot\nabla_{\bm{k}})^{2}\epsilon_{m}^{\prime}f_{D}^{\prime\prime\prime}(\epsilon_{m}^{\prime})+(e^2/2)\bm{E}\cdot\bm{g}_m\cdot\bm{E}f_D^{\prime\prime}(\epsilon^\prime_m)$.
Similarly, in the quantum RTA for the Liouville equation, Eq.~(\ref{eq:qrta}), the corresponding  quasiequilibrium density matrix $\rho^{\text{qe}}_{\bm{k}}$ should be taken as
\begin{align}
 \rho^{\text{qe}}_{\bm{k}}=  & \sum_m|u^\prime_{m\bm{k}}\rangle\langle u^\prime_{m\bm{k}}|\Biggl[f_D(\epsilon^\prime_{m}) -\frac{e^2}{24}(\bm{E}\cdot\nabla_{\bm{k}})^2\epsilon^\prime_m
f_D^{\prime\prime\prime}(\epsilon^\prime_m)\nonumber\\&+\frac{e^{2}}{2}\bm{E}\cdot\bm{g}_{m}\cdot\bm{E}f_D^{\prime\prime}(\epsilon^\prime_{m})\Biggr],
\end{align}
which yields Eq.~(\ref{eq:rho-metric}) when substituted into Eq.~(\ref{eq:rho-qrta}). 
This can be viewed as a generalization of the dynamical-phase approximation~\cite{Terada2024,Terada2025}, in which the field modulation of the dissipation term due to the dynamical phase factor is incorporated as a correction to the RTA.

\section{Concluding remarks}\label{sec:conclusion}
In summary, we have established a fully quantum formulation of nonlinear DC transport that can be cast as the following two equations, 
\begin{gather}
\bm{V}_{n}(\bm{k})  =\nabla_{\bm{k}}\epsilon^\prime_{n}(\bm{k}) +e\bm{E}\times\left(\nabla_{\bm{k}}\times\bm{a}^\prime_{n}(\bm{k}) \right),\\
-e\bm{E}\cdot\nabla_{\bm{k}}f_{n}(\bm{k}) =-\frac{f_{n}(\bm{k}) -f_{n}^{\text{qe}}(\bm{k}) }{\tau},
\end{gather}
The electric current is then given by $\bm{J}=-e\sum_n(2\pi)^{-3}\int d^3\bm{k}\bm{V}_n(\bm{k})f_n(\bm{k})$.
This formulation has the same structure as the semiclassical framework employed in Ref.~\cite{Sodemann2015}, but includes field-induced corrections to the eigenenergy $\epsilon^\prime_{n}(\bm{k})$, the Berry connection $\bm{a}^\prime_{n}(\bm{k})$, and the quasiequilibrium distribution function $f_{n}^{\text{qe}}(\bm{k})$. This framework has been shown to be valid to arbitrary order in $E$.
The discrepancies among previous studies stem from an incomplete treatment of these field-induced corrections, particularly those to the distribution function. Specifically, the full expression Eq.~(\ref{eq:qe-2}),
\begin{align*}
f_{n}^{\text{qe}}(\bm{k}) & =f_{D}(\epsilon_{n}^{\prime}(\bm{k}))-\frac{e^{2}}{24}(\bm{E}\cdot\nabla_{\bm{k}})^{2}\epsilon_{n}^{\prime}(\bm{k})f_{D}^{\prime\prime\prime}(\epsilon_{n}^{\prime}(\bm{k}))\\
&+\frac{e^{2}}{2}\bm{E}\cdot\bm{g}_{n}(\bm{k})\cdot\bm{E}f_{D}^{\prime\prime}(\epsilon_{n}^{\prime}(\bm{k}))+\dots,
\end{align*}
includes the long-overlooked quantum-metric contribution, as well as the eigenenergy correction proposed in semiclassical phenomenology. We have also clarified that the metric correction arises from positional fluctuations of the Bloch wave packet during relaxation under a bias field.

The quantum-metric correction to the distribution function unveiled in this study represents a novel type of geometric correction that cannot be absorbed into an eigenenergy correction; Otherwise, it could not contribute to the current response at
second order, as in the correction of Berry-connection-polarizability origin. 
It involves the nonequilibrium relaxation process and is absent when the coupling
to the reservoir (or more generally the collision term) is switched off, 
even though it is independent of the relaxation time $\tau$. 
Indeed, the lesser Green function Eq.~(\ref{eq:lesser}) describes the steady state and therefore implicitly assumes $t-t_0\gg\tau$, where $t_0$ denotes the initial time from which the relaxation proceeds.
At finite times, the correction should be accompanied by a transient factor such as $1-e^{-(t-t_0)/\tau}$,
which vanishes in the clean limit $\tau\to\infty$ at fixed $t-t_0$ and approaches unity only for $t-t_0\gg\tau$. The $\tau$-independent contribution here thus relies on taking the steady-state limit before the clean limit.
The above argument implies that a $\tau$-independent contribution should not necessarily be identified as intrinsic, unlike in the case of the linear Hall current: the notion of an intrinsic nonlinear response requires careful consideration, as also recognized in Ref.~\cite{Guo2026}. Following the original definition in linear response, a truly intrinsic current should be independent of the relaxation mechanism; in particular, it should survive even in clean systems.
The present metric contribution is not intrinsic in this sense.
This is also consistent with the necessity of dissipation to balance Joule heating in steady states.

Moreover, the quantum-metric term here indeed depends on the details of the relaxation mechanism, although it is independent of the dissipation strength $\Gamma$.
The direct manifestation of the quantum metric here relies on the
relaxation time being constant for all states. In particular, the lesser component of the self-energy, $2i\Gamma f_D(\omega)$
or $2i\Gamma f_D(\omega-e\bm{E}\cdot\hat{\bm{R}})$ for the two gauge
choices, is proportional to the identity operator on the internal
degrees of freedom. 
A matrix-valued $\Gamma$ in a more general scattering should modify the overlap that gives rise to the quantum metric in Eq.~(\ref{eq:metric-origin}) as
\begin{align}
\langle u_{n\bm{k}(t)}|\Gamma|u_{n\bm{k}(t^\prime)}\rangle&=\frac{1}{2}\sum_m \langle u_{n\bm{k}(t)}|\Gamma|u_{m\bm{k}(t)}\rangle\langle u_{m\bm{k}(t)}|u_{n\bm{k}(t^\prime)}\rangle\nonumber\\&
+\frac{1}{2}\sum_m\langle u_{n\bm{k}(t)}|u_{m\bm{k}(t^\prime)}\rangle\langle u_{m\bm{k}(t^\prime)}|\Gamma|u_{n\bm{k}(t^\prime)}\rangle.
\end{align}
Accordingly, the correction term should generally involve interband geometric quantities.

Despite the dependence on the relaxation mechanism, the involvement of quantum geometry itself is expected to be robust. 
Our result for the nonlinear current is equivalent to that of Ref.~\cite{Ulrich2026}, down to the exact
coefficients and signs (at zero temperature), in which the calculation is based
on the Matsubara Green function with a constant imaginary self-energy.
This agreement suggests that our result is not specific to the idealized reservoir model employed here.
While the relaxation time is taken to be constant throughout our calculation, identical lifetimes across the entire spectrum are not required in practice. 
Since the DC response is governed by a limited window of states near the Fermi surface, the present result should capture the leading contribution when the self-energy varies only weakly within this window. Its detailed momentum and frequency dependences then produce quantitative modifications without eliminating the geometric origin of the correction.
Also, as we have clarified through the length-gauge
picture, the quasiequilibrium distribution in the presence of an
electric field is subject to a strong constraint: it must cancel the
gauge dependence of the eigenenergy. This highlights the importance of
geometric quantities in dictating the quasiequilibrium distribution
that ensures this cancellation. 

Finally, although this study has focused on the nonlinear electric current, the quantum-metric correction should enter any observable, because it modifies the distribution function rather than the current operator.
In particular, while the nonreciprocal current of the quantum-metric origin requires breaking of both spatial-inversion and time-reversal symmetries, the quantum-metric correction to the observables of the other symmetry classes should be accessible in wider systems.

\begin{acknowledgments}
We thank Naoto Nagaosa for inspiring discussions. This work was supported by MEXT/JSPS KAKENHI, Grants Numbers JP25H01249, JP25K07219 (SK), JP26KJ0822 (TA), JP24H02231,
JP23K17665, JP24K00568 (TM), JP23K25816 (TM, SK) and JST SPRING, Grant Number JPMJSP2108 (TA).

\end{acknowledgments}

\appendix

\section{Derivation of the field-modulated adiabatic wave function\label{sec:appendix-derivation-C}}

In this appendix, we present the detailed derivation of the expression for $\tilde{C}_{nm}(\bm{k})$ [Eq.~(\ref{eq:C})],
starting from its equation of motion Eq.~(\ref{eq:eom-C}). By integrating Eq.~(\ref{eq:eom-C})
over $[-\infty,t]$ and performing a recursive substitution, we obtain
\begin{align}
C_{nm}(t) & =C_{nm}(-\infty)+i\int_{-\infty}^{t}dt_{1}\dot{\gamma}_{n\bm{k}_{0}}^{(2)}(t_{1})C_{nm}(t_{1})\nonumber \\
 & -ie\sum_{l\neq n}\int_{-\infty}^{t}dt_{1}\bm{E}(t_{1})\cdot\bm{\xi}_{nl}(\bm{k}(t_{1}))e^{i\gamma^\prime_{nl}(t_{1})}C_{lm}(t_{1}) \nonumber \\
 & =\delta_{nm}+i\int_{-\infty}^{t}dt_{1}\dot{\gamma}_{n\bm{k}_{0}}^{(2)}(t_{1})\delta_{nm}\nonumber \\
 & -ie\int_{-\infty}^{t}dt_{1}\bm{E}(t_{1})\cdot\bm{\xi}_{nm}(\bm{k}(t_{1}))e^{i\gamma^\prime_{nm}(t_{1})}(1-\delta_{nm})\nonumber \\
 & -e^{2}\sum_{l\neq n,m}\int_{-\infty}^{t}dt_{1}\bm{E}(t_{1})\cdot\bm{\xi}_{nl}(\bm{k}(t_{1}))e^{i\gamma^\prime_{nl}(t_{1})}\nonumber \\
 & \times\int_{-\infty}^{t_{1}}dt_{2}\bm{E}(t_{2})\cdot\bm{\xi}_{lm}(\bm{k}(t_{2}))e^{i\gamma^\prime_{lm}(t_{2})}+O(E^{3}).
\end{align}

Since the integral for the offdiagonal matrix element is oscillatory
because of the phase factor $\gamma^\prime_{nm}$, we can evaluate its
asymptotic form by a recursive use of integration by parts. Here
we assume an adiabatically-introduced electric field of the form
$\bm{E}(t)=\bm{E}e^{\eta t}$ with $\eta>0$.
We note that this procedure is employed merely as a technical device to obtain a solution independent of the initial profile of the electric field while preserving the norm. The resulting wave function need not be physically prepared by the same protocol.
We can then invoke the following relation
\begin{equation}
\bm{E}(t)e^{i\gamma_{nm}^{(0)}(t)}=-i\frac{1}{\Delta_{nm}(\bm{k}(t))-i\eta}\frac{d}{dt}\left[\bm{E}(t)e^{i\gamma_{nm}^{(0)}(t)}\right]
\end{equation}
for integration by parts. Applying this relation once for each oscillatory
integral and setting $\bm{E}(-\infty)=\bm{E}e^{\eta\times(-\infty)}=0$,
we obtain
\begin{widetext}
\begin{align}
C_{nm}(t) & =\delta_{nm}+i\int_{-\infty}^{t}dt_{1}\dot{\gamma}_{n\bm{k}_{0}}^{(2)}(t_{1})\delta_{nm}
-e\frac{\bm{E}(t)\cdot\bm{\xi}_{nm}(\bm{k}(t))}{\Delta_{nm}(\bm{k}(t))-i\eta}e^{i\gamma^\prime_{nm}(t)}(1-\delta_{nm})\nonumber \\
 & +e\int_{-\infty}^{t}dt_{1}e^{i\gamma^\prime_{nm}(t_{1})}\bm{E}(t_{1})
\cdot\Biggl[\left(\frac{d}{dt_{1}}+i\dot{\gamma}_{nm}^{(1)}(t_{1})\right)\left(\frac{\bm{\xi}_{nm}(\bm{k}(t_{1}))}{\Delta_{nm}(\bm{k}(t_{1}))-i\eta}\right)(1-\delta_{nm})
+ie\sum_{l\neq n,m}\frac{\bm{\xi}_{nl}(\bm{k}(t_{1}))\bm{E}(t_{1})\cdot\bm{\xi}_{lm}(\bm{k}(t_{1}))}{\Delta_{lm}(\bm{k}(t_{1}))-i\eta}\Biggr]+O(E^{3})\label{eq:C-osc1}
\end{align}
Noting that the $n=m$ contribution of the last line is not oscillatory,
we arrive at
\begin{align}
C_{nm}(t) & =\delta_{nm}+i\int_{t_{0}}^{t}dt_{1}\left[\dot{\gamma}_{n\bm{k}_{0}}^{(2)}(t_{1})+e^{2}\sum_{l\neq n}\frac{|\bm{E}(t_{1})\cdot\bm{\xi}_{nl}(\bm{k}(t_{1}))|^{2}}{\Delta_{ln}(\bm{k}(t_{1}))-i\eta}\right]\delta_{nm}\nonumber \\
 & +\Biggl[-\frac{e\bm{E}\cdot\bm{\xi}_{nm}(\bm{k}(t))}{\Delta_{nm}(\bm{k}(t))}+\frac{ie^{2}}{\Delta_{nm}(\bm{k}(t))}\bm{E}\cdot\hat{\bm{D}}_{nm}(\bm{k}(t))\left(\frac{\bm{E}\cdot\bm{\xi}_{nm}(\bm{k}(t))}{\Delta_{nm}(\bm{k}(t))}\right)+\sum_{l\neq n,m}\frac{e^{2}\bm{E}\cdot\bm{\xi}_{nl}(\bm{k}(t))\bm{E}\cdot\bm{\xi}_{lm}(\bm{k}(t))}{\Delta_{lm}(\bm{k}(t))\Delta_{nm}(\bm{k}(t))}\Biggr]\nonumber \\
 & \times e^{i\gamma^\prime_{nm}(t)}(1-\delta_{nm})+O(E^{3}),
\end{align}
\end{widetext}by performing integration by parts once again.
Here we take $\eta\to+0$ for the terms without an integral, assuming
no degeneracy in the Hamiltonian. Particular care must be taken with
the infinitesimal imaginary part in the non-oscillatory integral.
While the imaginary part of the non-oscillatory integral can be eliminated
by choosing 
\begin{equation}
\dot{\gamma}_{n\bm{k}_{0}}^{(2)}(t)=-e^{2}\sum_{l\neq n}\frac{|\bm{E}(t)\cdot\bm{\xi}_{nl}(\bm{k}(t))|^{2}}{\Delta_{ln}(\bm{k}(t))},
\end{equation}
the remaining part survives in the $\eta\to+0$ limit as
\begin{align}
 & i\int_{t_{0}}^{t}dt_{1}e^{2}\sum_{l\neq n}\frac{i\eta e^{2\eta t_{1}}|\bm{E}\cdot\bm{\xi}_{nl}(\bm{k}(t_{1}))|^{2}}{\Delta_{ln}(\bm{k}(t_{1}))(\Delta_{ln}(\bm{k}(t_{1}))-i\eta)}\nonumber \\
 & =-e^{2}\sum_{l\neq n}\frac{|\bm{E}(t)\cdot\bm{\xi}_{nl}(\bm{k}(t))|^{2}}{2\Delta_{ln}(\bm{k}(t))(\Delta_{ln}(\bm{k}(t))-i\eta)}+O(E^{3}).
\end{align}
This contribution keeps the norm of the wave function unity up to
$E^{2}$. Taking $\eta\to+0$, we arrive at Eq.~(\ref{eq:C}).

\section{Calculation of the overlap matrix element\label{sec:appendix-overlap}}

In this appendix, we derive the expression for the overlap matrix
element $\langle\psi^\prime_{n\bm{k}_{0}}(t-\tau)|\psi^\prime_{m\bm{k}_{0}}(t-\tau^{\prime})\rangle$ up to $E^2$.
In order to carry out the time convolution in Eq.~(\ref{eq:lesser}) for ${G}^{<}$, 
we perform the term-by-term integration through the series expansion of the wave
function $|\psi^\prime_{m\bm{k}_{0}}(t-\tau)\rangle$ with respect to $\tau$.
Note that all orders in $\tau$ will be retained below, while the $E^3$ contributions will be dropped.
First, we expand the dynamical phase factor $\gamma^{(d)}=\gamma^{(0)}-\gamma^{(2)}$ as
\begin{align}
e^{-i\gamma_{m\bm{k}_{0}}^{(d)}(t-\tau)}&=Q_{m}(\bm{k}(t),\tau)e^{-i\gamma_{m\bm{k}_{0}}^{(d)}(t)+i\epsilon_{m}(\bm{k}(t))\tau},\\
Q_{m}(\bm{k},\tau) & =1-i\sum_{n\neq m}\frac{|e\bm{E}\cdot\bm{\xi}_{mn}|^{2}}{\Delta_{mn}}\tau+\frac{ie}{2}\bm{E}\cdot\nabla_{\bm{k}}\epsilon_{m}\tau^{2}\nonumber \\
 & +\frac{ie^{2}}{6}(\bm{E}\cdot\nabla_{\bm{k}})^{2}\epsilon_{m}\tau^{3}-\frac{e^{2}}{8}(\bm{E}\cdot\nabla_{\bm{k}}\epsilon_{m})^{2}\tau^{4}.
\end{align}
Here we exclude the Berry-phase factor $\gamma^{(B)}=\gamma^{(1)}+2\gamma^{(2)}$ to keep the gauge invariance
of the expression. Through Eq.~(\ref{eq:freq}), the phase factor $\exp(i\epsilon_{m}(\bm{k}(t))\tau)$
here results in the expression with respect to $f_D(\epsilon_{m}(\bm{k}(t)))$
in the final form Eq.~(\ref{eq:f}), 
while the prefactor $Q$ describes the drift of the Fermi surface. 
Second, let us expand the modulated instantaneous eigenvector with the Berry
phase factor. Using a similar relation as Eq.~(\ref{eq:xi}), we obtain
\begin{align}
 |u_{m\bm{k}(t-\tau)}^{\prime}\rangle e^{-i\gamma_{m\bm{k}_{0}}^{(B)}(t-\tau)} & =|u_{m\bm{k}(t)}^{\prime}\rangle e^{-i\gamma_{m\bm{k}_{0}}^{(B)}(t)}\left(1-\frac{e^{2}}{2}\bm{E}\cdot\bm{g}_{m}\cdot\bm{E}\tau^{2}\right)\nonumber \\
 & +\sum_{n\neq m}|u_{n\bm{k}(t)}^{\prime}\rangle e^{-i\gamma_{m\bm{k}_{0}}^{(B)}(t)}\Biggl[-ie\tau\bm{E}\cdot\bm{\xi}_{nm}^{\prime}\nonumber \\
 &  -\frac{e^{2}\tau^{2}}{2}\langle u_{n\bm{k}(t)}|[\bm{E}\cdot(i\nabla_{\bm{k}}-\bm{a}_{m})]^{2}|u_{m\bm{k}(t)}\rangle\Biggr].
 \end{align}

Combining these contributions, we arrive at
\begin{align}
& |\psi_{m\bm{k}_{0}}^{\prime}(t-\tau)\rangle e^{i\gamma_{m\bm{k}_{0}}^{\prime}(t)-i\epsilon_{m}(\bm{k}(t))\tau}\nonumber \\
 & =|u_{m\bm{k}(t)}^{\prime}\rangle \Biggl[1-i\sum_{n\neq m}\frac{|e\bm{E}\cdot\bm{\xi}_{mn}|^{2}}{\Delta_{mn}}\tau+\frac{ie}{2}\bm{E}\cdot\nabla_{\bm{k}}\epsilon_{m}\tau^{2}\nonumber \\
 & -\frac{e^{2}}{2}\bm{E}\cdot\bm{g}_{m}\cdot\bm{E}\tau^{2}+\frac{ie^{2}}{6}(\bm{E}\cdot\nabla_{\bm{k}})^{2}\epsilon_{m}\tau^{3}-\frac{e^{2}}{8}(\bm{E}\cdot\nabla_{\bm{k}}\epsilon_{m})^{2}\tau^{4}\Biggr] \nonumber \\
 & +\sum_{n\neq m}|u_{n\bm{k}(t)}^{\prime}\rangle \Biggl[-ie\tau\bm{E}\cdot\bm{\xi}_{nm}^{\prime}\left(1+\frac{ie}{2}\bm{E}\cdot\nabla_{\bm{k}}\epsilon_{m}\tau^{2}\right)\nonumber \\
 & \qquad\qquad\quad -\frac{e^{2}\tau^{2}}{2}\langle u_{n\bm{k}(t)}|[\bm{E}\cdot(i\nabla_{\bm{k}}-\bm{a}_{m})]^{2}|u_{m\bm{k}(t)}\rangle\Biggr],
\end{align}
with which the diagonal part of the overlap matrix
is given by Eq.~(\ref{eq:S}).

We now confirm that the interband contribution to the electric current
should vanish after dropping $O(\Gamma)$ terms. The offdiagonal elements
of the overlap matrix are calculated as
\begin{align}
 & \langle\psi^\prime_{n\bm{k}_{0}}(t-\tau)|\psi^\prime_{m\bm{k}_{0}}(t-\tau^{\prime})\rangle =e^{i\gamma^\prime_{nm}(t)-i\epsilon_{n}(\bm{k}(t))\tau+i\epsilon_{m}(\bm{k}(t))\tau^{\prime}}\nonumber \\
  & \times\Biggl[ie\bm{E}\cdot\bm{\xi}_{nm}^{\prime}(\tau-\tau^{\prime})-\frac{e^{2}}{2}\sum_{l\neq n,m}\bm{E}\cdot\bm{\xi}_{nl}\bm{E}\cdot\bm{\xi}_{lm}(\tau-\tau^{\prime})^{2}\nonumber \\
 & +\frac{ie^{2}}{2}\bm{E}\cdot\hat{\bm{D}}_{nm}(\bm{E}\cdot\bm{\xi}_{nm})(\tau^{2}-\tau^{\prime2})\nonumber \\
 & +\frac{e^{2}}{2}\bm{E}\cdot\bm{\xi}_{nm}\left(\bm{E}\cdot\nabla_{\bm{k}}\epsilon_{n}\tau^{2}-\bm{E}\cdot\nabla_{\bm{k}}\epsilon_{m}\tau^{\prime2}\right)(\tau-\tau^{\prime})\Biggr].
\end{align}
Since the exponents for $\tau$ and $\tau^{\prime}$ are different,
we need to use the generalized version of the integral formula Eq.~(\ref{eq:freq}),
\begin{align}
 & 2\Gamma\int_{0}^{\infty}d\tau\int_{0}^{\infty}d\tau^{\prime}\int\frac{d\omega}{2\pi}f_D(\omega)(\tau+\tau^{\prime})^{n}(\tau-\tau^{\prime})^{m}\nonumber \\
 & \times e^{i(\omega-x)(\tau-\tau^{\prime})-(\Gamma-i\Delta)(\tau+\tau^{\prime})}=\frac{\Gamma n!}{(\Gamma-i\Delta)^{n+1}}i^{m}\frac{\partial^{m}f_D(x)}{\partial x^{m}}+O\left(\frac{\Gamma}{k_BT}\right).
\end{align}
Since $\Delta$ is nonnegligible compared with $\Gamma$ here, we
conclude that $[{G}_{\bm{k}_{0}}^{<}(t,t^{\prime})]_{nm}=O(\Gamma)$ for
$n\neq m$.

\section{Derivation of the quantum Boltzmann equation in the length gauge\label{sec:appendix-justification}}

In this appendix, we derive the quantum Boltzmann equation based on the length-gauge formalism. 
In particular, we show that the quantum Boltzmann equation is characterized by the quasiequilibrium distribution with the quantum correction of the form Eq.~(\ref{eq:intuitive}).

Before proceeding to the derivation, let us highlight the nontrivial role of the position operator in the length-gauge formalism by
explicitly constructing the solution of the time-dependent Schr\"odinger equation through the gauge transformation of Eq.~(\ref{eq:psi}). To this end, we introduce
the real-space representation of the velocity-gauge wave function,
\begin{align}
|\Psi_{m\bm{k}_{0}}(t)\rangle & =|\psi^\prime_{m\bm{k}_{0}}(t)\rangle\otimes|\bm{k}_{0}\rangle\nonumber \\
 & =|\psi^\prime_{m\bm{k}_{0}}(t)\rangle\otimes\left(\frac{1}{\sqrt{N}}\sum_{\bm{R}}|\bm{R}\rangle e^{i\bm{k}_{0}\cdot\bm{R}}\right),
\end{align}
where the wave function is expressed as a tensor product of the Hilbert
space for the internal degree of freedom [$\alpha$ in Eq.~(\ref{eq:htot})]
and the real space. The real-space part $\otimes|\bm{k}_{0}\rangle$
is omitted in the other sections, since all quantities in the velocity
gauge are diagonal in the momentum $\bm{k}_{0}$ because of translational
symmetry. In order to avoid the ill-definedness of the position operator, we take the number of sites $N\to\infty$, 
with which the momentum $\bm{k}$ becomes continuous. Although the Bloch wave becomes unnormalizable in $N\to\infty$, we employ the normalized expression assuming finite $N$ for the intermediate expressions for simplicity. Then the overlap of the real-space part is given by
$\langle\bm{k}|\bm{k}^{\prime}\rangle=(2\pi)^3N^{-1}\delta(\bm{k}-\bm{k}^{\prime})$, while the sum over $\bm{k}$-points is replaced into $\sum_{\bm{k}}=N\int_{\bm{k}}=N\int d^3\bm{k}/(2\pi)^3$. 

The length-gauge expression $|\Psi_{m\bm{k}_{0}}^{(\text{L})}(t)\rangle$
can be constructed via the gauge transformation $|\bm{R}\rangle\to|\bm{R}\rangle e^{ie\bm{A}(t)\cdot\bm{R}}$
as
\begin{align}
|\Psi_{m\bm{k}_{0}}^{(\text{L})}(t)\rangle & =|\psi^\prime_{m\bm{k}_{0}}(t)\rangle\otimes|\bm{k}(t)\rangle.\label{eq:psi-L}
\end{align}
This is indeed a solution of the time-dependent Schr\"{o}dinger
equation for the length-gauge Hamiltonian 
\begin{align}
\hat{H}^{(\text{L})} & =\hat{H}_{0}+e\bm{E}\cdot\hat{\bm{R}}\nonumber \\
 & =\sum_{\bm{k}}H(\bm{k})\otimes|\bm{k}\rangle\langle\bm{k}|+\sum_{\bm{R}}1\otimes e\bm{E}\cdot\bm{R}|\bm{R}\rangle\langle\bm{R}|,
\end{align}
which can be directly verified from
\begin{align}
 & i\partial_{t}|\Psi_{m\bm{k}_{0}}^{(\text{L})}(t)\rangle=\hat{H}^{(\text{L})}|\Psi_{m\bm{k}_{0}}^{(\text{L})}(t)\rangle\nonumber \\
 & =\frac{1}{\sqrt{N}}\sum_{\bm{R}}(H(\bm{k}(t))+e\bm{E}\cdot\bm{R})|\psi^\prime_{m\bm{k}_{0}}(t)\rangle\otimes|\bm{R}\rangle e^{i\bm{k}(t)\cdot\bm{R}},\label{eq:tdse-L}
\end{align}
using Eq.~(\ref{eq:tdse}) and $\dot{\bm{k}}(t)=-e\bm{E}$ for the evaluation of the time derivative.

We note that, although the gauge-transformed wave function Eq.~(\ref{eq:psi-L})
coincides with the ``eigenstate"
in the conventional length-gauge formalism $|u_{m\bm{k}}^{\prime}\rangle\otimes|\bm{k}\rangle$ as a snapshot at a fixed
$t$ up to the phase factor, its temporal dependence is not given by the standard form for
the static Hamiltonian, $|\Psi(t)\rangle=e^{-i\epsilon t}|\Psi(0)\rangle$.
This implies that the ``eigenstate"
is not an eigenstate of the length-gauge Hamiltonian, which is indeed 
the case due to the nontrivial action of the position operator. The action
of the position operator on the plane wave can be formally expressed
as $\hat{\bm{R}}|\bm{k}_{0}\rangle=-i\nabla_{\bm{k}}|\bm{k}_{0}\rangle$, 
with which one can show that the length-gauge eigenstate satisfies
\begin{align}
\hat{H}^{(\text{L})}|u_{m\bm{k}}^{\prime}\rangle\otimes|\bm{k}\rangle & =\epsilon_{m}^{\prime(\text{L})}(\bm{k})|u_{m\bm{k}}^{\prime}\rangle\otimes|\bm{k}\rangle-ie\bm{E}\cdot\nabla_{\bm{k}}(|u_{m\bm{k}}^{\prime}\rangle\otimes|\bm{k}\rangle).
\end{align}
The last term prevents the construction of the time-evolution operator through simple exponentiation, as it does not  
commute with the length-gauge eigenenergy
$\epsilon_{m}^{\prime(\text{L})}(\bm{k})=\epsilon_{m}^{\prime}(\bm{k})+e\bm{E}\cdot\bm{a}_{m}^{\prime}(\bm{k})$ [Eq.~(\ref{eq:eps-len})].

Accordingly, the explicit expressions for the nonequilibrium Green functions $\tilde{G}$ in the length gauge take a nontrivial form, 
\begin{align}
\tilde{G}^{R}(\omega) & =-i\sum_{n\bm{k}}\int_{0}^{\infty}dt e^{-i\int_{0}^{t}dt^\prime\epsilon_{n}^{\prime(\text{L})}(\bm{k}-e\bm{E}t^\prime)}\nonumber \\
 & \times e^{i\omega t-\Gamma t}|u_{n,\bm{k}-e\bm{E}t}^{\prime}\rangle\langle u_{n\bm{k}}^{\prime}|\otimes|\bm{k}-e\bm{E}t\rangle\langle\bm{k}|,\\
\tilde{G}^{<}(\omega) & =2i\Gamma\tilde{G}^{R}(\omega)f_{D}(\omega-e\bm{E}\cdot\hat{\bm{R}})\tilde{G}^{A}(\omega).
\end{align}
Although $\tilde{G}$ has time-translational symmetry and admits a frequency-space expression, it becomes nondiagonal in momentum as a drawback. 

In order to avoid this nontrivial temporal structure, in the length-gauge formalism, it is more convenient to employ the equation of motion for the Green functions (or the density matrix~\cite{Ventura2017}) rather than that for the wave function. For the matrix element
\begin{equation}
\tilde{G}_{nm}^{R}(\bm{k},\bm{k}^{\prime},\omega)=(\langle u_{n\bm{k}}^{\prime}|\otimes\langle\bm{k}|)\,\tilde{G}^{R}(\omega)\,(|u_{m\bm{k}^{\prime}}^{\prime}\rangle\otimes|\bm{k}^{\prime}\rangle),
\end{equation}
the equation of motion reads
\begin{multline}
[\omega-\epsilon_{n}^{\prime(\text{L})}(\bm{k})+i\Gamma]\tilde{G}_{nm}^{R}(\bm{k},\bm{k}^{\prime},\omega)\\
=\frac{(2\pi)^{3}}{N}\delta(\bm{k}-\bm{k}^{\prime})\delta_{nm}+ie\bm{E}\cdot\nabla_{\bm{k}}\tilde{G}_{nm}^{R}(\bm{k},\bm{k}^{\prime},\omega).\label{eq:dyson-R-len}
\end{multline}
Again, the last term can be traced back to the nontrivial action of the position operator.
While Eq.~(\ref{eq:dyson-R-len}) can be viewed as the Dyson equation with $\tilde{G}^R_0=[\omega-\epsilon_{n}^{\prime(\text{L})}(\bm{k})+i\Gamma]^{-1}$ and $\tilde{\Sigma}^R=ie\bm{E}\cdot\nabla_{\bm{k}}$, it is inappropriate to drop the self-energy $\tilde{\Sigma}^R$ to yield the leading-order expression of the standard form $\tilde{G}^R=\tilde{G}^R_0$, because the full expression involves the derivative of the delta function $(\bm{E}\cdot\nabla_{\bm{k}})^n\delta(\bm{k}-\bm{k}^{\prime})$, which is closely related to the indefiniteness of the position for the plane waves and should not be dropped. One can circumvent this singular form by considering the equation of motion for $\int_{\bm{k}^\prime}e^{i(\bm{k}-\bm{k}^\prime)\cdot\bm{R}_0}\tilde{G}_{nm}^{R}(\bm{k},\bm{k}^{\prime},\omega)$ instead.
The leading-order expression for $\tilde{G}_{nm}^{R}(\bm{k},\bm{k}^{\prime},\omega)$ is then obtained by the inverse transform as
\begin{equation}
\tilde{G}_{nm}^{R}(\bm{k},\bm{k}^{\prime},\omega)\sim\frac{1}{N}\sum_{\bm{R}_{0}}\frac{\delta_{nm}e^{-i(\bm{k}-\bm{k}^{\prime})\cdot\bm{R}_{0}}}{\omega-\epsilon_{n}^{\prime}(\bm{k})-e\bm{E}\cdot[\bm{R}_{0}+\bm{a}_{n}^{\prime}(\bm{k})]+i\Gamma}.\label{eq:GR-len}
\end{equation}
Note that this expression is nondiagonal in momentum and deviates from the standard form for the static Hamiltonian.

In order to take advantage of the length gauge, we also consider the equation of motion for the lesser Green function, rather than its explicit form shown above. This is nothing but the quantum Boltzmann equation. In particular, the nonequilibrium distribution function defined with the diagonal component,
\begin{align}
f_{n\bm{k}} & =-i\int\frac{d\omega}{2\pi}(\langle u_{n\bm{k}}^{\prime}|\otimes\langle\bm{k}|)\,\tilde{G}^{<}(\omega)\,(|u_{n\bm{k}}^{\prime}\rangle\otimes|\bm{k}\rangle),
\end{align}
satisfies the equation of motion of the semiclassical Boltzmann form,
$-e\bm{E}\cdot\nabla_{\bm{k}}f_{n\bm{k}}=-2\Gamma(f_{n\bm{k}}-f_{n\bm{k}}^{\text{qe}})$ [Eq.~(\ref{eq:rta})].
Here the quasiequilibrium distribution $f_{n\bm{k}}^{\text{qe}}$ to relax is given by
\begin{align}
f_{n\bm{k}}^{\text{qe}} & =\text{Re}\sum_{m\bm{k}^{\prime}}\int\frac{d\omega}{2\pi}2i\tilde{G}_{nm}^{R}(\bm{k},\bm{k}^{\prime},\omega)\nonumber \\
 & \times(\langle u_{m\bm{k}^{\prime}}^{\prime}|\otimes\langle\bm{k}^{\prime}|)\,f_{D}(\omega-e\bm{E}\cdot\hat{\bm{R}})\,(|u_{n\bm{k}}^{\prime}\rangle\otimes|\bm{k}\rangle).
\end{align}
By adopting the leading-order expression Eq.~(\ref{eq:GR-len}) for the retarded component and taking its pole for the frequency integral, the quasiequilibrium distribution can be approximated as
\begin{multline}
f_{n\bm{k}}^{\text{qe}}\sim\frac{1}{N}\text{Re}\sum_{\bm{k}^{\prime}\bm{R}_{0}}e^{i(\bm{k}-\bm{k}^{\prime})\cdot\bm{R}_{0}}(\langle u_{n,\bm{k}}^{\prime}|\otimes\langle\bm{k}|)\\
\times f_{D}(\epsilon_{n}^{\prime}(\bm{k})-e\bm{E}\cdot[\hat{\bm{R}}-\bm{R}_{0}-\bm{a}_{n}^{\prime}(\bm{k})])(|u_{n\bm{k}^{\prime}}^{\prime}\rangle\otimes|\bm{k}^{\prime}\rangle).
\end{multline}
This expression can be identified with Eq.~(\ref{eq:intuitive}) by introducing the site-resolved expectation value by
\begin{align}
\langle\hat{O}\rangle_{n\bm{k};\bm{R}_{0}} & =\text{Re}\sum_{\bm{k}^{\prime}}e^{i(\bm{k}-\bm{k}^{\prime})\cdot\bm{R}_{0}}(\langle u_{n\bm{k}}^{\prime}|\otimes\langle\bm{k}|)\,\hat{O}\,(|u_{n\bm{k}^{\prime}}^{\prime}\rangle\otimes|\bm{k}^{\prime}\rangle)\nonumber \\
 & =\text{Re}\sum_{\bm{R}_{0}^{\prime}}\langle w_{n\bm{R}_{0}^{\prime}}|\hat{O}|w_{n\bm{R}_{0}}\rangle e^{i\bm{k}\cdot(\bm{R}_{0}-\bm{R}_{0}^{\prime})},
\end{align}
where $|w_{n\bm{R}_{0}}\rangle=N^{-1/2}\sum_{\bm{k}}e^{-i\bm{k}\cdot\bm{R}_{0}}(|u_{n\bm{k}}^{\prime}\rangle\otimes|\bm{k}\rangle)$ is the Wannier orbital at site $\bm{R}_0$ (with hybrid Wannier orbitals used if necessary).
In particular, the action of the position operator on the plane wave can be transferred to the internal degree of freedom through integration by parts, as
$\int_{\bm{k}^{\prime}}F(\bm{k}^{\prime})|u_{n\bm{k}^{\prime}}^{\prime}\rangle\otimes(\hat{\bm{R}}|\bm{k}^{\prime}\rangle)=\int_{\bm{k}^{\prime}}i\nabla_{\bm{k}^{\prime}}[F(\bm{k}^{\prime})|u_{n\bm{k}^{\prime}}^{\prime}\rangle]\otimes|\bm{k}^{\prime}\rangle$. This yields a simplified expression Eq.~(\ref{eq:site-resolved-R}).

\section{Relaxation-time equation with higher-order terms}\label{sec:appendix-higher}

In this appendix, we derive a formal expression for the quasiequilibrium distribution function that generalizes Eq.~(\ref{eq:intuitive}) to arbitrary order in $E$.
To this end, we first express the nonequilibrium distribution function $f_{m,\bm{k}(t)}=-i[G_{\bm{k}_{0}}^{<}(t,t)]_{mm}$ obtained from the lesser Green function Eq.~(\ref{eq:lesser}) as
\begin{multline}
f_{m,\bm{k}(t)} =-i\int_{0}^{\infty}d\tau\int_{0}^{\infty}d\tau^{\prime}e^{-\Gamma(\tau+\tau^{\prime})}\\
\times e^{-\frac{\tau+\tau^{\prime}}{2}\partial_{t}}\left[\Sigma_{\bm{k}_{0}}^{<}\left(t-\frac{\tau-\tau^{\prime}}{2},t+\frac{\tau-\tau^{\prime}}{2}\right)\right]_{mm}.
\end{multline}
Since the wave-function overlap in the self-energy $\Sigma^<$ [Eq.~(\ref{eq:sigma-lesser})] can be rewritten as
\begin{align}
\langle\psi_{m\bm{k}_{0}}^{\prime}(t-t^{\prime})|\psi_{m\bm{k}_{0}}^{\prime}(t+t^{\prime})\rangle & =\langle\psi_{m,\bm{k}(t)}^{\prime}(-t^{\prime})|\psi_{m,\bm{k}(t)}^{\prime}(t^{\prime})\rangle
\end{align}
with $\bm{k}(t)=\bm{k}_0-e\bm{E}t$, we can replace time derivatives with momentum derivatives as $\partial_{t}^N\Sigma_{\bm{k}_{0}}^{<}(t-t^{\prime},t+t^{\prime})=(-e\bm{E}\cdot\nabla_{\bm{k}})^N\Sigma_{\bm{k}(t)}^{<}(-t^{\prime},t^{\prime})$.
The integral formula Eq.~(\ref{eq:freq}) then gives
\begin{gather}
f_{m\bm{k}}=\sum_{N=0}^{\infty}\left(\frac{e}{2\Gamma}\bm{E}\cdot\nabla_{\bm{k}}\right)^{N}f_{m\bm{k}}^{\text{qe}},\\
f_{m\bm{k}}^{\text{qe}}  =-i\int_{0}^{\infty}d\tau\int_{0}^{\infty}d\tau^{\prime}e^{-\Gamma(\tau+\tau^{\prime})}\left[\Sigma_{\bm{k}}^{<}\left(-\frac{\tau-\tau^{\prime}}{2},\frac{\tau-\tau^{\prime}}{2}\right)\right]_{mm}.
\end{gather}
This expression shows that the nonequilibrium distribution function satisfies the relaxation-time equation, with $f^{\text{qe}}$ as the quasiequilibrium distribution toward which it relaxes:
\begin{align}
-e\bm{E}\cdot\nabla_{\bm{k}}f_{m\bm{k}} & =-2\Gamma(f_{m\bm{k}}-f_{m\bm{k}}^{\text{qe}}).
\end{align}
The series expansion of the dynamical phase $\gamma^{(d)}$ in $t=(\tau-\tau^\prime)/2$,
\begin{equation}
\gamma_{m\bm{k}}^{(d)}(t)-\gamma_{m\bm{k}}^{(d)}(-t)=\sum_{N=0}^{\infty}\frac{2t^{2N+1}}{(2N+1)!}\left(e\bm{E}\cdot\nabla_{\bm{k}}\right)^{2N}\epsilon_{m}^{\prime}(\bm{k}),
\end{equation}
allows us to evaluate the integral in $f_{m\bm{k}}^{\text{qe}}$ by means of term-by-term integration with respect to $E$. 
Separating the $N=0$ contribution $2\epsilon_m^\prime t$ from the dynamical phase and applying Eq.~(\ref{eq:freq}) to the remaining part, we obtain
\begin{align}
f_{m\bm{k}}^{\text{qe}} & =\sum_{M=0}^{\infty}\frac{i^{M}}{2^MM!}\frac{\partial^{M}f_{D}(\epsilon_{m}^{\prime})}{\partial\epsilon^{M}}\left.\frac{\partial^{M}}{\partial t^{ M}}[\langle\psi_{m\bm{k}}^{\prime}(-t)|\psi_{m\bm{k}}^{\prime}(t)\rangle e^{2i\epsilon_{m}^{\prime}t}]\right|_{t=0}.
\end{align}
This expression reproduces both the quantum-metric contribution and the third-derivative term at second order in $E$, as shown in Eq.~(\ref{eq:qe-2}).

\bibliography{references}

\end{document}